\documentclass[letter, traditabstract]{aa} % for the letters 
\usepackage{graphicx}
\usepackage[varg]{txfonts}
\usepackage{upgreek} 
\usepackage{bigdelim}
\usepackage{multirow} %para combinar lineas en las tablas
\usepackage[FIGTOPCAP, center, nooneline]{subfigure}
\usepackage{natbib}
\usepackage{rotating}
\usepackage{lscape}
\usepackage{float}

\bibpunct{(}{)}{;}{a}{}{,}

\usepackage{setspace} %para variar el interlineado

\usepackage{color} %Para poner palabras en color
\begin{document}

   \title{New molecular species at redshift $z=0.89$
\thanks{This paper is based on observations with the 40\,m radio telescope
 of the National Geographic Institute of Spain (IGN) at Yebes Observatory.
Yebes Observatory thanks the ERC for funding support under grant ERC-2013-Syg-610256-NANOCOSMOS.}}

     \author{B. Tercero\inst{\ref{inst1},\ref{inst2}} \and J. Cernicharo\inst{\ref{inst3}} 
     \and S. Cuadrado\inst{\ref{inst3}} \and P. de Vicente\inst{\ref{inst2}} \and M. Gu\'elin\inst{\ref{inst4}}}

   \institute{Observatorio Astron\'omico Nacional (OAN-IGN). Calle Alfonso XII, 3, E-28014 Madrid, Spain.\label{inst1}
   \and Observatorio de Yebes (IGN). Cerro de la Palera s/n, E-19141 Yebes, Guadalajara, Spain.\label{inst2}
   \and Instituto de F\'{\i}sica Fundamental (IFF-CSIC). Calle Serrano 123, E-28006 Madrid, Spain. \label{inst3}
   \and Institut de Radioastronomie Millim\'etrique, 300 rue de la Piscine, 38406 Saint Martin d'H\`eres, France.\label{inst4}
   \\
\email{b.tercero@oan.es}}

   \date{Received - 2020; accepted - 2020}

% \abstract{}{}{}{}{} 
% 5 {} token are mandatory
 
  \abstract{
  %\LEt{ I edited your paper to UK spelling and grammar conventions.}
  We present the first detections of CH$_3$SH,
C$_3$H$^+$, C$_3$N, HCOOH, CH$_2$CHCN, and H$_2$CN in an extragalactic
source. Namely the spiral arm of a galaxy located
at $z=0.89$ on the line of sight to the radio-loud quasar PKS 1830$-$211.
OCS, SO$_2$, and NH$_2$CN were also detected, 
raising the total number of molecular species identified in that early time
galaxy to 54, not counting isotopologues. The
detections were made in absorption against the SW quasar image, at 2
kpc from the galaxy centre, over the course of a Q band 
spectral line survey made with the Yebes 40\,m telescope (rest-frame
frequencies: 58.7\,$-$\,93.5\,GHz). 
%\LEt{ A\&A uses the past tense to describe
%specific methods used in a paper, and the present tense to describe general
%methods and the findings of recent papers. See Sect. 4 of the language guide
%https://www.aanda.org/for-authors/language-editing/4-verb-tense-and-voice.\ Please review my edits to ensure the appropriate changes were made. }
We derived the rotational temperatures
and column densities of those species, which are found to be
subthermally excited. The molecular abundances, and in particular the large 
abundances of C$_3$H$^+$ and of several previously reported cations, 
are characteristic of diffuse or translucent clouds with enhanced UV 
radiation or strong shocks.}

  % context heading (optional)
  % {} leave it empty if necessary  
  % aims heading (mandatory)
  % methods heading (mandatory) 
  % results heading (mandatory)
   
  % conclusions heading (optional), leave it empty if necessary 


   \maketitle
%
%________________________________________________________________
%**********************************************************************
\section{Introduction}\label{Intro}

Some 220 molecular species, not counting isotopologues, have been identified in the
galactic interstellar medium (ISM). Among them are ions, neutral radicals, metallic
compounds, and complex organic molecules (COMs), whose relative abundances and isotopic ratios 
vary drastically due to the type of source and environment
(see CDMS\footnote{\texttt{https://cdms.astro.uni-koeln.de/}}). Conversely,
the observation of molecular abundances offers a powerful way to
characterise the gas properties and past history. This is particularly
true for distant galaxies where we lack spatial resolution.

Distant galaxies tell us about the earlier stages of the ISM.
Heavy elements and metals appeared early on in the ISM
after a first generation of massive stars released their
nucleosynthesis products into space. Secondary elements, such as
nitrogen, mostly appeared several
billions of years later; after a second generation of stars,
low-mass longer-lived stars released their processed material 
\citep{Johnson2019}. The chemical and isotopic
composition of the ISM in a galaxy at redshift $z=1$, when the Universe
was half its present age, should therefore be markedly different from
that in the Galaxy and be dominated by the products of massive stars
\citep{Muller2011}. An analysis of the molecular content and isotopic
ratios in distant galaxies, thus, offers us an opportunity to characterise
these types of sources and to check current models of galaxy chemical
evolution.
%In addition, the detections of molecules at intermediate and high redshifts
%have served as interesting cosmological probes (see \citealt{Muller2014a} and references therein).

A powerful method to study the ISM of distant galaxies is through the
detection of molecular lines in absorption.  The strength of the
absorption depends on the flux of the background source and the line
opacity. 
The latter can be directly obtained and, if the background
quasar image is small enough to be fully covered by the absorbing gas,
this strength is decoupled from the distance to the absorber. Therefore, 
it is not diluted by the telescope beam,
allowing for the detection of low abundance species
in distant sources.

%Then, this strength is diluted neither by the distance to the absorber, nor
%by the telescope beam, allowing the detection of low abundance species
%in distant sources.

One of the most studied molecular absorbers is the system located
towards the blazar PKS\,1830$-$211 at redshift $z=2.5$
(\citealt{Lidman1999}; see \citealt{Muller2006} for a detailed
description of the system). The blazar image is gravitationally lensed
by a foreground, nearly face-on spiral galaxy at $z=0.89$ that
intercepts the line of sight \citep{Wiklind1996,Winn2002}. At radio wavelengths, lensing
gives rise to two point-like images of the blazar,
embedded in a faint Einstein ring of 1$''$ in diameter (8\,kpc at the
distance of the galaxy, \citealt{Jauncey1991}). The two bright images,
that is, only the ones that are still present at millimetre wavelengths, are
located 2\,kpc SW and 6\,kpc NE of the galaxy nucleus, respectively. The redshift of
$z$\,=\,0.89 corresponds to an age of 6.4\,Gyr (see \citealt{Wright2006}
for the assumed cosmological parameters) and hence to a lookback time
close to half the present age of the Universe.

The molecular absorption arises in two spiral arms symmetrically
unrolling on either side of the nucleus \citep{Wiklind1996}.  Several
studies using millimetre interferometers (NOEMA, ATCA, and ALMA) have
revealed the molecular richness of the SW arm \citep{Muller2006,
Muller2011, Muller2013, Muller2014a, Muller2014b, Muller2016a,
Muller2016b, Muller2017, Muller2015}. More than 45
molecular species, plus more than 15 rare isotopologues, have been
reported in those studies. Among others, they consist of COMs 
(such as CH$_3$OH, CH$_3$CN, NH$_2$CHO, and
CH$_3$CHO), of hydrocarbons (l-C$_3$H, l-C$_3$H$_2$, and C$_4$H), and of
light hydrides and cations (H$_2$Cl$^+$, ArH$^+$, CF$^+$, OH$^+$,
H$_2$O$^+$, CH$^+$, and SH$^+$). \citet{Muller2011, Muller2014a}
suggest that this denotes a chemical signature similar to that of the
galactic diffuse and 
%\LEt{ Please avoid the use of the slash. For details, refer to Sect. 3.6 of the A\&A Author's Guide.}
translucent clouds. Furthermore, the analysis of the
light hydrides detected in this source points out a multi-phase
composition of the absorbing gas \citep{Muller2016a, Muller2016b}.

PKS\,1830$-$211 is also a strong gamma-ray emitter, with several flaring
events in recent years (see e.g. \citealt{Abdo2015}).
During 2019, flares of intense gamma-ray activity were reported  
using the Large Area Telescope (LAT), one of the two instruments on board the 
Fermi Gamma-ray Space Telescope \citep{Buson2019}. The preliminary analysis conducted by 
the Fermi-LAT collaboration indicates that this source
has been undergoing a long-term brightening since October 2018.
Following this gamma-ray outburst, a radio-wave monitoring campaign with the 32\,m Medicina and 64\,m Sardinia
radio telescopes, at 8.3\,GHz and 25.4\,GHz, has revealed a parallel radio emission 
enhancement \citep{Iacolina2019}. 

We took advantage of this event to search for new extragalactic
molecular species with the Yebes 40\,m radio telescope, whose 7\,mm
receiver was recently updated with the help of the European Research
Council NANOCOSMOS project\footnote{\texttt{https://nanocosmos.iff.csic.es/}}.  
In this Letter,
we report the detection of six new extragalactic molecular species towards 
the SW image of PKS\,1830$-$211.  We derived their molecular column densities
and rotational temperatures, as well as those of three more
species that were detected for the first time in this source
(Section\,\ref{results}). In light of the new
identifications, we discuss the implications for the prevailing
chemistry in the source (Section\,\ref{dis}).

\section{Observations and data reduction}
\label{obs}

The observations began in April 2019 with the 40\,m diameter Yebes
radio telescope\footnote{\texttt{http://rt40m.oan.es/rt40m$\_$en.php}}, just after
the recent enhancement of the blazar radio flux \citep{Buson2019,Iacolina2019}.
They consisted in monitoring the continuum
flux and absorption line profiles in the course of commissioning
the new 7\,mm receiver (F. Tercero et al. in preparation). The
telescope \citep{deVicente2016} is located at 990\,m of altitude near
Guadalajara, Spain.

The Q band (7\,mm) HEMT receiver allows for simultaneous broad-band
observations in two linear polarisations. It is connected to 16 fast
Fourier transform spectrometers (FFTS), each covering 2.5\,GHz with a
38\,kHz channel separation. In each polarisation, this system provides an instantaneous band of 18\,GHz between 31.5\,GHz and 50\,GHz.

The 7\,mm spectra shown in this paper are the average of 12 observing
sessions, which were carried out between April 2019 and September 2019.  Due to
the low declination of the source, it could only be observed for 5 hours per day
above 15$^{\circ}$ elevation (the telescope latitude is
$+$40$^{\circ}$\,31$'$\,29.6$''$). The intensity scale was calibrated
using two absorbers at different temperatures and the atmospheric
transmission model (ATM, \citealt{Cernicharo1985, Pardo2001}).

In addition, we carried out three observing sessions at 1.3\,cm
with the K band receiver on 
%\LEt{ Cardinal ending are not allowed on dates at A\&A because it is informal style.}
May 29 and 30 as well as June 2.  The goal of those observations was to confirm the
identification of C$_3$H$^+$ by observing a third rotational
transition. The covered band extended from 23.7\,GHz to 24.2\,GHz,
and the channel spacing was 30\,kHz.  
For these data, the intensity
scale was calibrated using a noise diode and the ATM model.

The observational procedure was position switching with the
reference position located $-$240$''$ away in azimuth. The telescope
pointing and focus were checked every one to two hours through
pseudo-continuum observations of VX\,Sgr, a red hypergiant star close
to the target source. VX\,Sgr shows strong SiO $v$\,=\,1
$J$\,=\,1$-$0 (at 43.122\,GHz) and H$_2$O
$J_{K_a,K_c}$\,=\,6$_{1,6}$$-$5$_{2,3}$ (at 22.235\,GHz) maser
emission.
Pseudo-continuum observations consist in subtracting
the summed spectral emission from the masers from the rest
of the spectrum, while performing pointing azimuth-elevation cross drifts or a focus
scan. This is a useful technique since atmospheric emissivity fluctuations during the scans
have a small influence on the results and provide very flat continuum baselines.
In addition, we regularly checked the pointing towards the blazar
PKS\,1830$-$211 itself (coordinates \mbox{$\upalpha_{\rm
J2000}$\,=\,18$^{\rm h}$\,33$^{\rm m}$\,39.9$^{\rm s}$},
\mbox{$\updelta_{\rm J2000}$\,=\,$-$21$^{\circ}$\,03$'$\,39.6$''$}),
whose flux (NE\,+\,SW images) was $\simeq$\,10\,Jy at intermediate frequencies of the 7\,mm band.
%by adapting the operations of the FFT spectrometers to continuum observations 
%by averaging the whole channels of the FFT spectrometers
%during the pointing drifts.
The pointing errors were always found $<5''$ on both axes.  In order
to confirm that no spurious signals were contaminating our spectra,
the frequency of the local oscillator was changed by 20\,MHz 
from one observing session to the next.

The data were reduced using the \texttt{CLASS} software of the
\texttt{GILDAS}
package\footnote{\texttt{http://www.iram.fr/IRAMFR/GILDAS/}}.  A
polynomial baseline of low order (typically second or third order) was
removed from the spectrum obtained for each session. All sessions were then
averaged together. In this paper, we adopt $z$\,=\,0.885875
($v_0$\,=\,0\,km\,s$^{-1}$ in the local standard of rest, LSR) to
convert the observed frequencies to rest-frame frequencies. 
%The final spectra are in units of antenna
%temperature, $T_A^*$, corrected for atmospheric absorption and
%spillover losses. 
For each observed frequency, Table\,\ref{table_telescope} lists the
telescope aperture efficiency ($\eta_{\rm A}$), the conversion factor
between flux ($S$) and antenna temperature ($T_{\rm A}^*$), the half power
beam width (HPBW), the system temperature ($T_{\rm sys}$), the integration
time, and the noise root mean square (rms).
%\LEt{ Single-sentence paragraphs are not allowed.}

\section{Continuum emission}
\label{cont}

The intensity scales of the spectra shown in this Letter were
normalised to the total (NE\,+\,SW) continuum intensity of the source. The latter
varies with frequency as indicated (in $T_A^*$ scale) in
Table\,\ref{table_telescope}. The molecular lines presented in this
article all appear in absorption against the SW image (the NE image
absorption, which is much weaker, appears at markedly different
velocities).  In order to derive the line opacities, we had to
determine the contribution of the SW image to the total continuum flux
detected by the telescope. This can be done by assuming that the
HCO$^+$ and HCN ($J=1-0$) line absorption (see Fig.\,\ref{fig_hcn_hco})
is at its total at velocities near $v_0=0$\,km\,s$^{-1}$, which corresponds to
the arm located in front of the SW source. This assumption, which was already
discussed by \citet{Muller2008} for the $J=2-1$ lines, is supported by
the obvious saturation of the flattened absorption profiles and by the
near-stability of the saturation level despite strong blazar intensity
variations.

\citet{Muller2008} measured a roughly constant flux ratio of the NE to
SW components, $R$\,$\sim$\,1.7, over several observations between
1995 and 2007 at 3\,mm. This value is similar to that obtained by
\citet{Subrahmanyan1990} at the centimetre wavelengths. Using ATCA data from 2011, \citet{Muller2013} derived an $R$ of $\sim$1.5 at 3\,mm and
7\,mm.  Following the strategy of \citet{Muller2008}, we estimated the
NE/SW flux ratio using the  $J=1-0$ HCO$^+$ and HCN
lines. As these saturated lines
absorb 45\,\% of the total flux, the NE component (with the
residual emission of the faint Einstein ring) has to contribute 55\,\%
to the total flux. We obtain a rather constant value of
$R$\,$\sim$\,1.2 over the period from April to September 2019 (see
Fig.\,\ref{fig_hcn_hco}).  This value is close to that measured by
\citet{Muller2014a} at 250\,GHz and 290\,GHz from ALMA data from April 9 and 11, 2012, respectively, just before a
gamma-ray flare. It is, however, significantly lower than in
most of the other measurements at millimetre and submillimetre
wavelengths (see also \citealt{Marti2019}).\ This is perhaps due to a conjunction
between the blazar flux variations and the four-week delay between the 
NE and SW light paths. These flux variations are thought to be
associated with plasmon ejections along the precessing jet of the
blazar \citep{Nair2005,Muller2008}.

\begin{figure*}%[!ht]
\includegraphics[scale=0.65,angle=0]{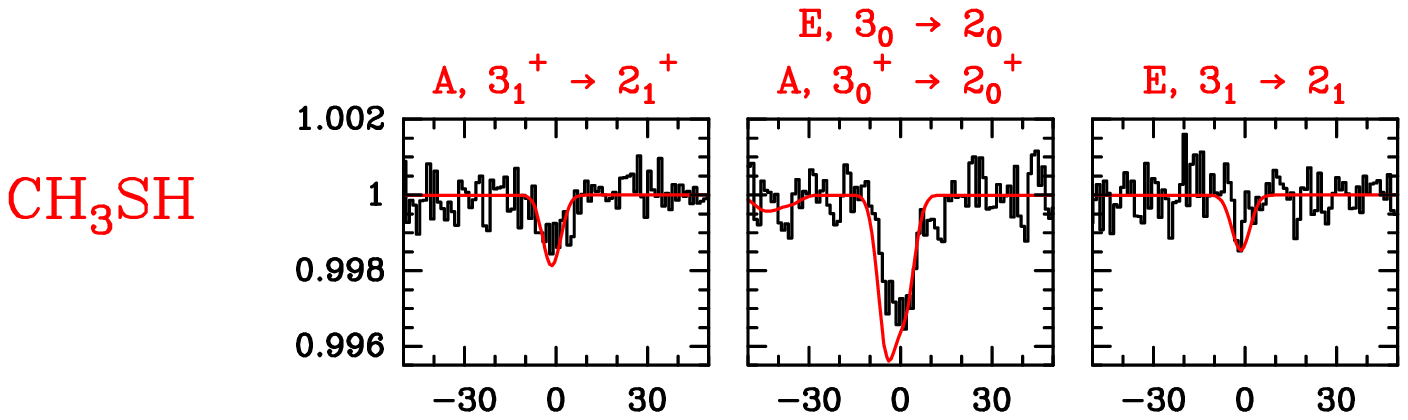}  %\vspace*{0.1cm}

\includegraphics[scale=0.65,angle=0]{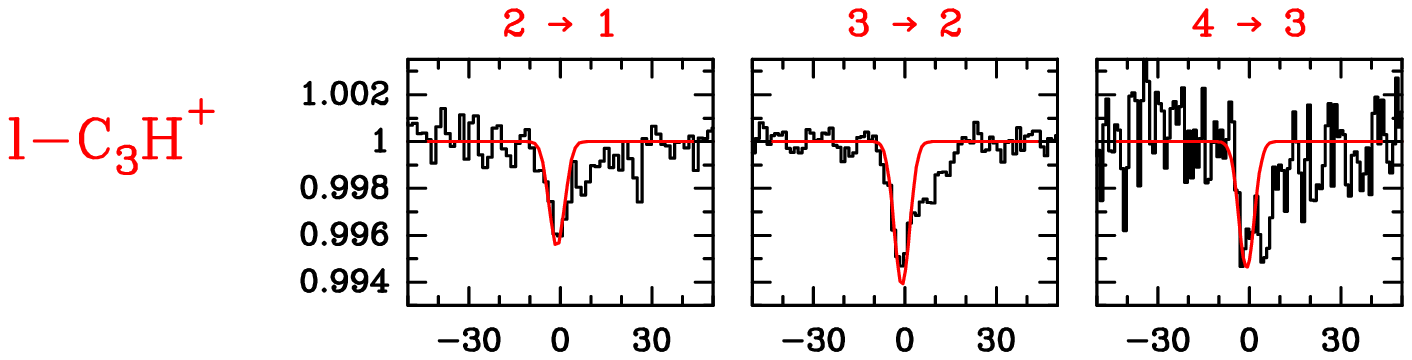}  %\vspace*{0.05cm}

\includegraphics[scale=0.65,angle=0]{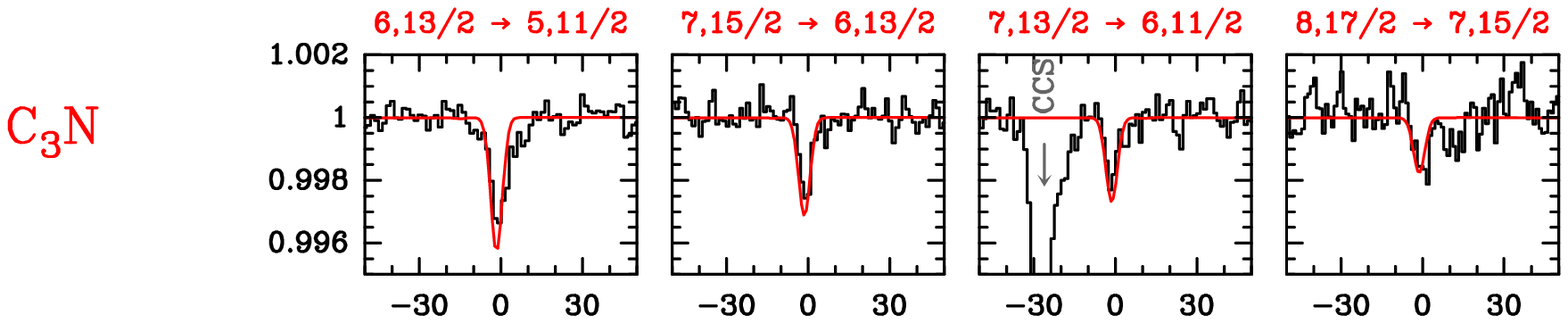}  %\vspace*{0.05cm}

\includegraphics[scale=0.65,angle=0]{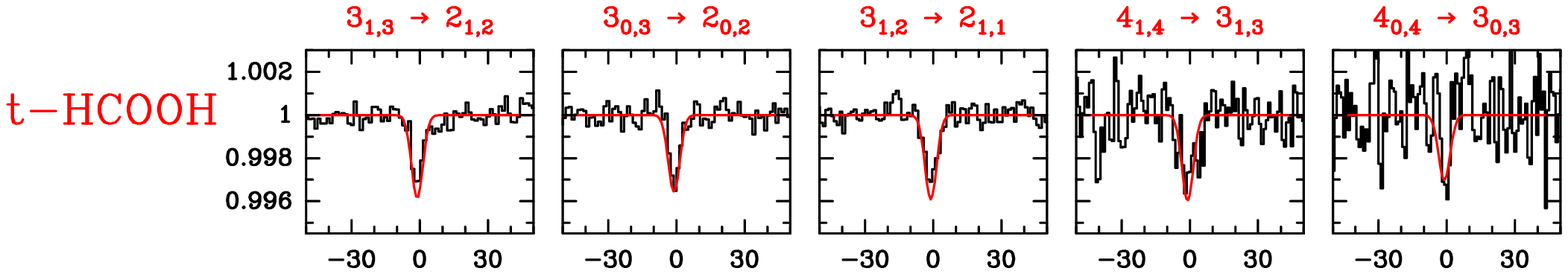}  %\vspace*{0.1cm}

\includegraphics[scale=0.65,angle=0]{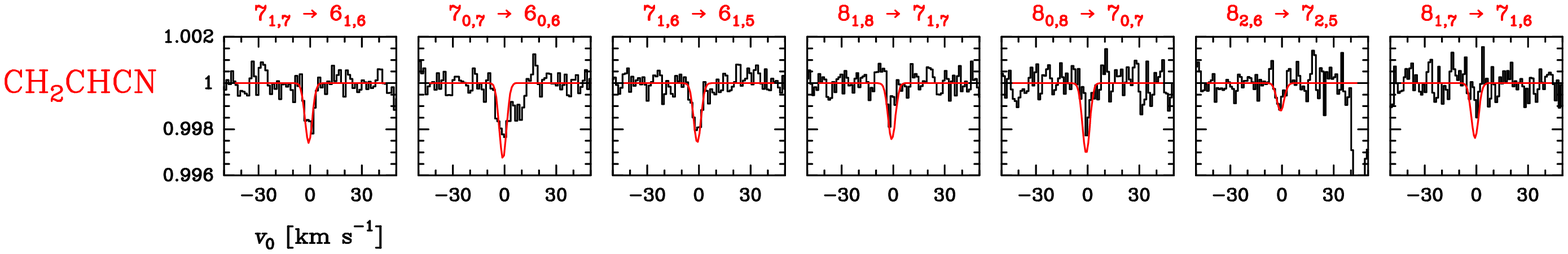}  %\vspace*{0.2cm}

\includegraphics[scale=0.65,angle=0]{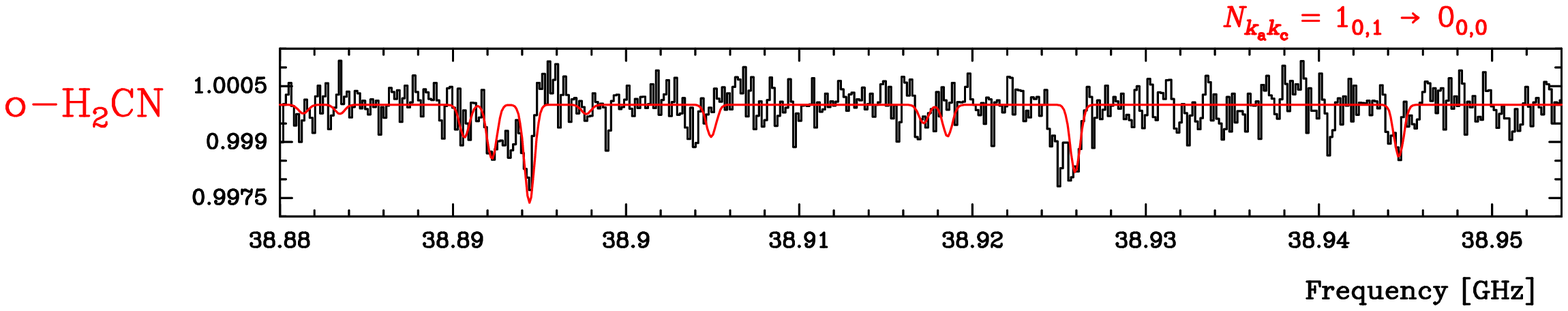} 

\includegraphics[scale=0.65,angle=0]{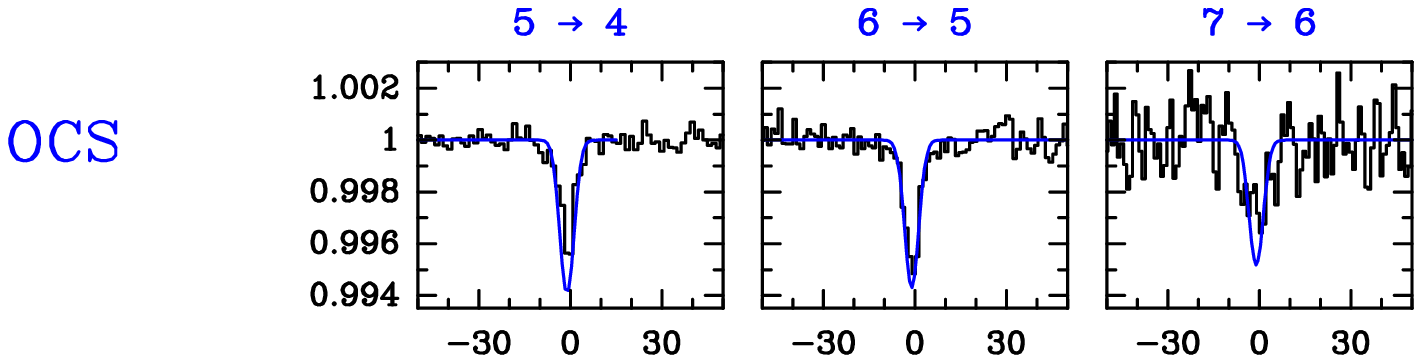}  %\vspace*{0.1cm}

\includegraphics[scale=0.65,angle=0]{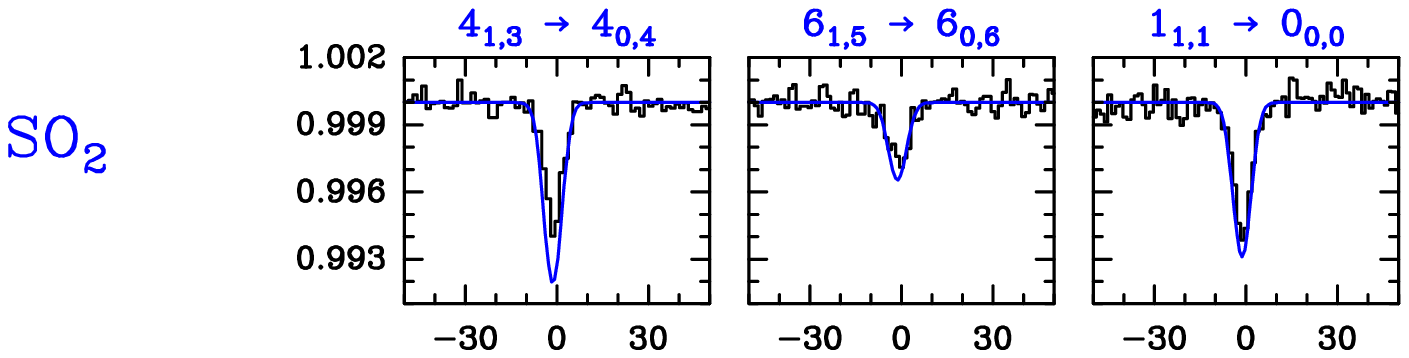}  %\vspace*{0.1cm}

\includegraphics[scale=0.65,angle=0]{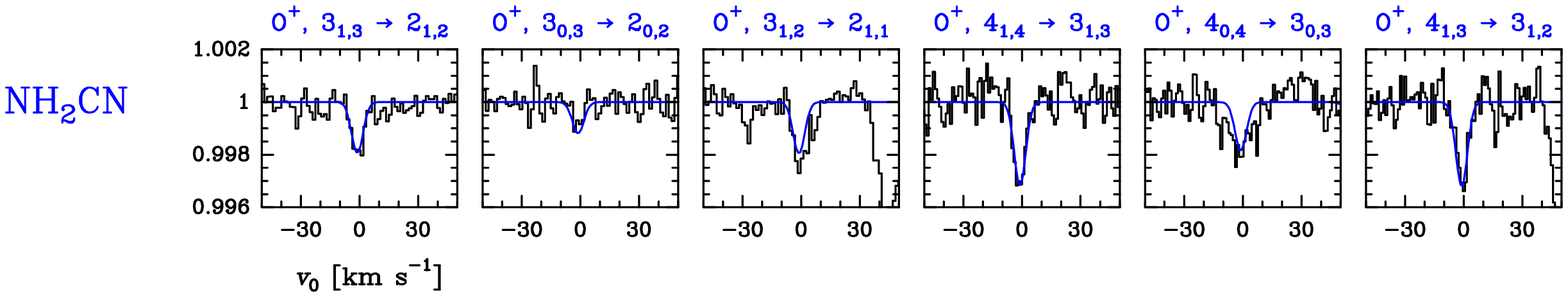} %\vspace*{0.1cm}
\caption{Absorption lines towards PKS\,1830$-$211 observed with the Yebes 40\,m telescope. 
The species identified for the first time in an extragalactic source are highlighted in red.
The molecules highlighted in blue are those that were identified for the first time in this source, but observed previously in other extragalactic sources.
The $y$-axis is the normalised intensity to the total (NE\,+\,SW) continuum
level (see Sect.\,\ref{cont}).
We assume a $z$\,=\,0.885875 ($v_{\rm 0}$\,=\,0\,km\,s$^{-1}$ in the local standard of rest).}
\label{fig_lines}
\end{figure*}

\section{Results}
\label{results}

\begin{table*}[!ht]
\centering
\caption{Dipole moments ($\mu$), number of detected lines, table number
listing the spectroscopic and observational line parameters, rotational temperatures ($T_{\rm rot}$) and column densities ($N$) obtained with the rotational diagram (RD) analysis, and
physical parameters assumed for the \texttt{MADEX} LTE model of the new molecules detected in this work.}
\label{Table_dipole_moments}
\begin{tabular}{l c c c c c c c c c c c@{\vrule height 8pt depth 4pt width 0pt}}
\hline
\hline
 & & &  & \multicolumn{2}{c}{RD analysis} \rule[0.2cm]{0cm}{0.15cm}\ & &  \multicolumn{4}{c}{\texttt{MADEX} LTE model} \rule[0.2cm]{0cm}{0.15cm}\ & \\ \cline{5-6} \cline{8-11}
 
Molecule  & $\mu$   &  No. Lines & Table & $T_{\rm rot}$ & $N$\,$\times$\,10$^{12}$ & & $v_{\rm 0}$ &  $v_{\rm FWHM}$ & $T_{\rm rot}$ & $N$\,$\times$\,10$^{12}$ \rule[0.2cm]{0cm}{0.15cm}\ & Notes \\
  & [Debye]  &   &  & [K] & [cm$^{-2}$] & & [km\,s$^{-1}$] &  [km\,s$^{-1}$] & [K] & [cm$^{-2}$]  & \\
\hline
A/E-CH$_{3}$SH      &   $\mu_{a}$=1.3$^{a}$     & 3   & \ref{Table_CH3SH}    & ...                & ...                  &   & $-$1.0 &   7.0 & 5.0   &  20 &  \\
l-C$_{3}$H$^{+}$    &   $\mu_{a}$=3.0$^{b}$     & 3   & \ref{Table_l-C3H+}   & 5\,$\pm$\,1        & 1.2\,$\pm$\,0.2      &   & $-$0.6 &   6.0 & 5.0   &  1.0 &  \\
C$_{3}$N            &   $\mu_{a}$=2.8$^{c}$     & 4   &  \ref{Table_C3N}     & 5\,$\pm$\,1        & 4\,$\pm$\,1       &   &   $-$0.5   &   5.0 & 5.0   &  \,3.0 & $\dagger$   \\
t-HCOOH             &   $\mu_{a}$=1.4$^{d}$     & 5   & \ref{Table_HCOOH}    & 5\,$\pm$\,1        & 2.3\,$\pm$\,0.5   &   & $-$0.7 &   6.0 & 5.0   &  2.3 & * \\
                    &                           &     &                      & 6\,$\pm$\,1        & 3.3\,$\pm$\,0.4   &   & $-$0.7 &   6.0 & 6.0     &  3.3 & **    \\
                    &                           &     &                      &                    & 6\,$\pm$\,2       &   &  &    &    &   &***  \\
CH$_{2}$CHCN        &   $\mu_{a}$=3.8$^{e}$     & 7   & \ref{Table_CH2CHCN}  & 10\,$\pm$\,4\,\,\, & 5\,$\pm$\,2          &   & $-$0.5 &   5.0 & 10    &  5.5 &  \\
o-H$_{2}$CN         &   $\mu_{a}$=2.5$^{f}$     & 4   & \ref{Table_o-H2CN}   & ...                & ...                  &   & $-$1.0 &   5.0 & 5.0   &  4.0  & \\
\hline                                                                                                                                                
OCS                 &   $\mu_{a}$=0.7$^{g}$     &  3  &  \ref{Table_OCS}     & 8\,$\pm$\,3        & 43\,$\pm$\,15        &   & $-$0.7 &   5.5 & 8.0     &  43  & \\
SO$_{2}$            &   $\mu_{b}$=1.6$^{h}$     &  3  &  \ref{Table_SO2}     & 8\,$\pm$\,1        & 25\,$\pm$\,4\,\,\,   &   & $-$0.9 &   7.0 & 8.0     &  25  &\\
O$^+$-NH$_{2}$CN          &   $\mu_{a}$=4.3$^{i}$     &  6  & \ref{Table_NH2CN}    & 30\,$\pm$\,5\,\,\, & 11\,$\pm$\,2\,\,\,   &   & $-$0.9 &   7.0 & 30    &  11 & \\
\hline

         \end{tabular}
         \tablefoot{Refs.
         (a)~\citet{Tsunekawa1989},
         (b)~\citet{Pety_2012},
         (c)~\citet{McCarthy1995},
         (d)~\citet{Kuze1982},
         (e)~\citet{Krasnicki2011},
         (f)~\citet{Cowles1991},
         (g)~\citet{Tanaka_1985},
         (h)~\citet{Patel_1979},
         (i)~\citet{Read1986}.\\
         ($\dagger$) This slight difference in $N$ (within the
uncertainty) between the two methods is mainly caused by adapting the C$_3$N spectroscopic
values in the rotational diagram in order to address the blending of
the hyperfine components (see\,Sect.\,\ref{appendix_dr}). (*) For the $K_{\rm a}=0$ ladder (see\,Sect.\,\ref{appendix_dr}). (**) For the $K_{\rm a}=1$ ladder (see\,Sect.\,\ref{appendix_dr}). 
(***) Total column density (see\,Sect.\,\ref{appendix_dr}).
}
   \end{table*}

%&&&&&&&&&&&&&&&&&&&&&&&&&&&&&

For the first time, we detected
CH$_3$SH, C$_3$H$^+$, C$_3$N, HCOOH, CH$_2$CHCN, and H$_2$CN in an extragalactic source and identified the following three more
species, also for the first time, towards PKS\,1830$-$211: OCS, SO$_2$, and NH$_2$CN. 
All of those species have been detected in absorption towards the SW spiral arm of the $z$\,=\,0.89 
galaxy in front of the blazar. Figure\,\ref{fig_lines} shows these detections. For each species, Table\,\ref{Table_dipole_moments}
lists its dipole moment and derived properties.

A single Gaussian profile was fitted to the strongest
component of all detected lines to obtain the observational line parameters: radial velocities ($v_{\rm 0}$), full width at half maximum (FWHM) line widths ($v_{\rm FWHM}$), 
and absorption intensities ($I_{\rm \nu}$) measured from the unity on the normalised spectrum to the total (SW\,+\,NW) continuum level (see Sect.\,\ref{cont}). Then,
optical depths ($\tau_{\rm \nu}$) towards the SW line of sight were directly derived according to

\begin{equation}
\tau_{\rm \nu} = -{\rm ln} \left(1 - \frac{I_{\rm \nu}}{0.45}\right)
\label{eq_tau}
,\end{equation}
assuming an SW source covering factor of unity (see e.g. \citealt{Muller2008} and \citealt{Muller2011}; these references
also address a more general discussion regarding how to obtain optical depths of absorbing lines in this source).
This assumption does not introduce major uncertainties since \citet{Muller2014a}, 
using ALMA observations, conclude that the size of the SW continuum image is only 5\,$-$\,10\,\%
larger than the size of the absorbing gas at millimetre and submillimetre wavelengths.
These results, together with the spectroscopic parameters of the involved transitions are shown in Tables\,\ref{Table_CH3SH}\,$-$\,\ref{Table_NH2CN}.
We note that while the derived radial velocities are in agreement with
previous observations at 7\,mm towards the SW line of sight \citep{Muller2011,Muller2013},
most of the line widths of
the molecular lines in our data are in a range
between 5\,$-$10\,km\,s$^{-1}$. Our absorbing molecular lines
are narrower than those of these previous works by a factor of 2.
Narrower lines
may indicate either lower optical depths or different clouds intercepting the
line of sight. This variation could result from
the appearance or disappearance of discrete components in the
background continuum source \citep{Muller2008}.

As we detected several rotational transitions for many species,
we estimated their rotational temperatures ($T_{\rm rot}$)
and molecular column densities ($N$) by drawing 
rotational diagrams.
%e.g. \citealt{Goldsmith1999}).
The analysis assumes the Rayleigh-Jeans approximation, optically thin lines,
and local thermodynamic equilibrium (LTE). In Appendix\,\ref{appendix_dr},
we describe this method and the diagrams relative to this work (Fig.\,\ref{fig_DR}). 
The resulting $T_{\rm rot}$ and $N$ are given in Table\,\ref{Table_dipole_moments}.

We note that the $T_{\rm rot}$ obtained for C$_3$H$^+$, C$_3$N, and HCOOH is close
to the temperature of the cosmic microwave background (CMB) at
$z$\,=\,0.89 ($T_{\rm CMB}$\,=\,5.08\,$\pm$\,0.10\,K,
\citealt{Muller2013}).  This result was already noted in previous
surveys
\citep{Combes1999,Menten1999,Henkel2009,Muller2011,Muller2013}.  It
implies an excitation dominated by radiative coupling with the CMB
since the gas kinetic temperature in the SW source is found from
NH$_3$ lines that are much higher ($\sim$80\,K, according to
\citealt{Henkel2008}).  It also means that the bulk of the gas has a
moderate or low density: $\leq 10^3$\,cm$^{-3}$ (e.g. \citealt{Henkel2009,Muller2013}).

It is important to note that CH$_2$CHCN, OCS, and SO$_2$ have a somewhat higher $T_{\rm rot}$
(8\,$-$\,10\,K).\ This suggests that either the observed lines arise in the denser parts
of the absorbing clouds or these molecules present 
lower critical densities than those of C$_3$H$^+$, C$_3$N, and HCOOH, which are, therefore, more sensitive to collisions.

The case is more extreme for NH$_2$CN for which we find $T_{\rm
rot}=30$\,K.  Interestingly enough, the lines pertaining to the two
different $K_{\rm a}$ ladders of that latter species merge into a
single straight line in the rotational diagram after correcting for the
different statistical weights. This is typical of asymmetric top
molecules in dense environments (see Appendix\,\ref{appendix_dr}).
As we mention above, these results may indicate that either
NH$_2$CN traces a denser and hotter component of the absorbing gas
or a surprisingly low critical density for this species.

To confirm the line identifications, we produced synthetic spectra of
the putative species using the \texttt{MADEX} tool
\citep{Cernicharo2012}.  For this, we assumed LTE approximation and
adopted the $T_{\rm rot}$ and $N$ values derived from the rotational
diagrams.  In the cases of CH$_3$SH and H$_2$CN, for which we do not have enough lines to draw a rotational diagram, we fixed $T_{\rm rot}$ to
5\,K and varied $N$ until we obtained a reasonable fit to the line profiles. The 
$T_{\rm rot}$ and $N$ values as well as the fitted $v_{\rm 0}$ and $v_{\rm FWHM}$ for all
species are shown in Table\,\ref{Table_dipole_moments}.  Since
\texttt{MADEX} computes the line opacities 
%and brightness temperatures ($T'_{\rm B}$) 
in the absence
of a continuum background source ($\tau_{\rm \nu}$), we derived the line absorption profile   
against the SW continuum image, normalised to the total (NE\,+\,SW) continuum emission by
re-arranging Eq.\,\ref{eq_tau} to obtain $I_{\rm \nu}$.
%through the
%relation:
%\begin{equation}
%I_{\rm \nu} = \left( \frac{T'_{\rm B} + (0.45 T'_{\rm cont}) {\rm e}^{-\tau_{\rm \nu}}}{T'_{\rm %cont}} \right) + 0.55
%\label{eq_model}
%\end{equation}
%where $T'_{\rm B}$ and $T'_{\rm cont}$ are 
%beam-averaged line and continuum brightness temperatures
%(assuming the absorbing clouds fully cover the SW continuum image, see e.g. \citealt{Muller2008}).
%which also depends on the frequency.
We note that the absorption by the SW spiral arm occurs at velocities of $0\pm 50$ km\,s$^{-1}$, which are quite different from those
caused by the NE arm around $-$140 km\,s$^{-1}$, and that no absorption is detected at the latter velocity for the nine 
species considered here. 
%In practice, $T'_{\rm B}$ is negligible compared with the term \mbox{(0.45 $T'_{\rm cont}$) %e$^{-\tau_{\rm \nu}}$} and $\tau$ is low so the relation take the
%form of Eq.\,\ref{eq_tau}.
%so that $\tau_{nu}$ is readily calculated from Iv (see Table\,\ref{table_telescope}).
%\textbf{***well, we should be careful here as the size of the arm is larger than that of the SW continuum image***}
%It is worth noting that the value of $T'_{\rm B}$ is affected by the dilution factor of the telescope beam.
%Therefore, as $T'_{\rm B}$\,$\rightarrow$\,0 
%and the spectrum is normalised to the total continuum level,
%so $T'_{\rm cont}$ can be considered in antenna temperature (see Table\,\ref{table_telescope}). 
The resulting synthetic spectrum is overlaid over the observed lines in Fig.\,\ref{fig_lines}.
This simple model reproduces the observed lines.
The model also allows us to distinguish an additional component for C$_3$H$^{+}$ and
H$_2$CN between 0\,km\,s$^{-1}$ and 10\,km\,s$^{-1}$ that was not taken into account in our model. This component is also
seen in the lowest energy lines of CH$_3$SH and CH$_2$CHCN.
However, the weakness of this component in the mentioned lines
prevents us from further evaluating its properties. 
                      
\section{Discussion}
\label{dis}
This work %largely 
expands the inventory of extragalactic species, particularly 
in the SW line of sight towards PKS\,1830$-$211 %at $v_{\rm 0}$\,$\sim$\,0\,km\,s$^{-1}$ 
with the detection of nine new species.
With these new species, the number of molecular species towards
the SW line of sight of PKS\,1830$-$211 raises to more than 50.

The sample of molecules detected here 
%complicates the interpretation of the nature of the absorbing gas and 
points to a large diversity in molecular environments:
\begin{enumerate}
\item
Carbon
chains (C$_3$N) and ion-molecule gas phase-produced $-$CN species
(H$_2$CN and CH$_2$CHCN) have been detected in dense
($\sim$10$^4$\,cm$^{-3}$) and cold ($T_{\rm K}$\,$\sim$\,10\,K)
prestellar cores \citep{Ohishi1994,Kaifu2004} of our Galaxy.
\item
CH$_3$SH and NH$_2$CN have only been detected towards hot cores
in high-mass star-forming regions
\citep{Turner1975,Linke1979,Kolesnikova2014} and low-mass protostars
\citep{Cernicharo2012b,Majumdar2016,Coutens2018}. %In addition,
NH$_2$CN has also been detected in the starburst galaxies M32 and NGC253 \citep{Martin2006,Aladro2011}.
Besides prestellar cores, CH$_2$CHCN is an abundant species in hot cores \citep{Belloche2013,Lopez2014}.
%On the other hand, extragalactic 
\item
Although OCS, SO$_2$, and HCOOH have been detected in a large variety of environments, these molecules usually trace
hot and shocked gas in star forming regions \citep{Tercero2010,Esplugues2013,Tercero2018} . 
\item
C$_3$H$^+$ traces 
the edge of dense molecular clouds illuminated by UV radiation. In our Galaxy,
this species has only been unambiguously detected towards the Horsehead and the Orion Bar \citep{Pety_2012,McGuire2014,Cuadrado2015},
which are two well known photodissociation regions (PDRs). 
After the first detection of C$_3$H$^+$ towards the Horsehead \citep{Pety_2012}, 
\citet{McGuire2013} tentatively detected C$_3$H$^+$ via absorption lines ($J$\,=1$-$0, 2$-$1) in diffuse,
spiral arm clouds along the line of sight to Sgr\,B2(N).
These authors \citep{McGuire2014}, using the CSO telescope
operating at 1\,mm, conducted a further search for C$_3$H$^+$ towards 39 galactic sources including 
hot cores, evolved stars, dark clouds, class 0 objects, PDRs, HII regions, outflows, and translucent clouds.
Interestingly, they only detected this species towards the Orion Bar, a prototypical
high-UV flux, hot PDR, with a far-UV radiation field of a few 10$^4$ times the mean interstellar
field (see \citealt{Cuadrado2015} and references therein). It is possible that
the negative results of \citet{McGuire2014} towards translucent clouds were due to a bias introduced by
the high energies of the searched transitions ($E_{\rm u}$/$k_{\rm B}$\,$>$\,50\,K). Although the gas in PDRs is also
subthermally excited, these regions are usually hotter and denser ($T_{\rm K}$\,$\sim$\,150\,K and $n$(H$_{2}$)\,$\sim$\,10$^5$\,cm$^{-3}$
for the Orion Bar, see \citealt{Cuadrado2015}) than translucent clouds,
allowing for the excitation of more energetic transitions.

\end{enumerate}

A large variety of conditions coexist in sources such as SgrB2, and absorption 
measurements against a bright and hot continuum source are a powerful way of revealing all types of clouds.
%\LEt{ Single-sentence paragraphs are not allowed.}
%Although it is diffucult to explain the presence of this variety of molecules in a single environment,
%this view of a diversity of molecular clouds responsible for the absorbing lines of the different species
%towards the SW line of sight collides with the uniform values for the observed line parameters of the detected species. 
%\textbf{***I wouldn't say this that way:
As we mention in Sect.\,\ref{results}, the gas intercepting the SW line of sight at $v_{\rm 0}$\,$\sim$\,0\,km\,s$^{-1}$ mostly
consists of translucent clouds with a moderate density (10$^3$\,cm\,$^{-3}$) and
with a kinetic temperature of $\sim$80\,K \citep{Henkel2009,Muller2013}.
The wide detection of cations towards the SW line of sight \citep{Muller2014b,Muller2015,Muller2016a,Muller2016b,Muller2017} indicates
that non-thermal mechanisms, such as high UV or X-ray illumination of the gas or shocks, dominate
the chemistry of this region.
The previous tentative detection of C$_3$H$^+$ in diffuse spiral arm clouds towards Sgr\,B2(N)
and its detection towards PKS\,1830$-$211
suggests that translucent clouds may produce this cation in significant abundances.
The gas-phase production of C$_3$H$^+$ via C$_2$H$_2$\,+\,C$^+$ 
competes with its destruction pathway via H$_2$ leading to small hydrocarbons \citep{Pety_2012}.
To observe detectable abundances of C$_3$H$^+$, the UV-radiation field should be greatly enhanced relative
to the mean interstellar value to produce sufficient C$^+$.
The detection of C$_3$H$^+$ towards the translucent clouds intercepting the SW line of sight of PKS\,1830$-$211
indicates the large average UV-field in this medium.
%The detection of C$_3$H$^+$ towards the translucent clouds intercepting the SW line of sight of PKS\,1830$-$211
%might also indicate a larger average UV-field 
%interestellar value 
%in distant galaxies.

It is worth noting that many COMs in our Galaxy have been detected, for the first time, 
in absorption and in the centimetre domain
pointing towards the hot cores of Sgr\,B2 (see e.g. CH$_2$CHCHO, CH$_3$CH$_2$CHO, CH$_2$OHCHO,
c-H$_2$C$_3$O, CH$_3$CONH$_2$, CH$_3$CHNHO, HNCHCN, and CH$_3$CHCH$_2$O;
\citealt{Hollis2004a,Hollis2004b,Hollis2006a,Hollis2006b,Loomis2013,Zaleski2013,McGuire2016}).
These detections via absorption lines indicate that these species are associated with
the moderate dense ($n$(H$_2$)\,$\sim$\,10$^3$\,$-$\,10$^4$\,cm$^{-3}$) 
and warm ($T_{\rm K}$\,$\simeq$\,100\,$-$\,300\,K) envelope of Sgr\,B2.
This envelope around the hot cores near the galactic centre also consists of a subthermally excited
molecular gas \citep{Huettemeister1995,deVicente1997,Jones2008,Etxaluze2013}. 
This suggests that both the heating of this envelope and the production of these complex species
might be mainly related to 
non-thermal processes, such as shocks
or an enhanced UV or X-ray flux in the surrounding medium.
Thus, we may consider that similar mechanisms may produce the diversity of
molecular species detected towards the translucent clouds in the SW line of
sight towards PKS\,1830$-$211.

%**********************************************************************

\begin{acknowledgements}
We would like to thank the anonymous referee for a helpful
report that led to improvements in the paper.
We thank the ERC for support under grant ERC-2013-Syg-610256- NANOCOSMOS. We also thank
the Spanish MINECO for funding support under grants AYA2012-32032 and FIS2014-52172-C2,
and  the CONSOLIDER-Ingenio programme $``$ASTROMOL$"$ CSD 2009-00038.
\end{acknowledgements}

%**********************************************************************

\bibliographystyle{aa}
\bibliography{references}
%Wright2006 PASP 118 1711

%**********************************************************************

\clearpage

\begin{appendix}

\section{Complementary material}

\begin{table*}[!ht]%t1
\begin{center}
\caption{Aperture efficiency ($\eta_{\rm A}$), antenna temperature ($T_{\rm A}^*$) to flux ($S$) conversion factor, half power beam width (HPBW), system temperature ($T_{\rm sys}$), integration time, noise root mean square (rms), and source continuum emission in $T_{\rm A}^*$ along the covered frequency range.}\label{table_telescope}
%\resizebox{1\textwidth}{!}{
  \begin{tabular}{cccccccccc@{\vrule height 8pt depth 4pt width 0pt}}
\hline
\hline
  Frequency & $\eta_{\rm A}$ & $S$/$T_{\rm A}^*$ & HPBW & $T_{\rm
sys}$$^{(a)}$ & $T_{\rm sys}$$^{(b)}$ [K] & Integration & rms & $T_{\rm
A}^*$$^{(a)}$ [K] & $T_{\rm A}^*$$^{(b)}$ [K] (cont.) \\
  $[$GHz]   &               &     [Jy/K]        & [$''$] & [K] & average &
time$^{(c)}$ [h] & [mK] & (continuum) & average \\
\hline
24.0 & 0.50 & 4.2 & 72.2 & 102\,$-$\,112 & 108 & 10.5 & 5.3 &
2.9\,$-$\,3.1 & 3.0 \\
32.5 & 0.45 & 4.5 & 53.5 & 73\,$-$\,124 & 100 & 35.7 & 2.4 &
2.3\,$-$\,3.0 & 2.7 \\
34.7 & 0.44 & 4.6 & 50.1 & 75\,$-$\,123 & 100 & 37.9 & 1.8 &
2.3\,$-$\,3.0 & 2.5 \\
37.0 & 0.43 & 4.7 & 47.0 & 83\,$-$\,132 & 107 & 35.7 & 1.7 &
2.0\,$-$\,2.7 & 2.4 \\
39.3 & 0.42 & 4.8 & 44.2 & 91\,$-$\,141 & 115 & 35.7 & 1.8 &
1.8\,$-$\,2.5 & 2.3 \\
41.6 & 0.41 & 4.9 & 41.7 & 112\,$-$\,161 & 135 & 35.7 & 2.3 &
1.6\,$-$\,2.4 & 2.2 \\
43.9 & 0.39 & 5.2 & 39.6 & 138\,$-$\,188 & 157 & 37.9 & 2.6 &
1.2\,$-$\,2.2 & 1.9 \\
46.2 & 0.37 & 5.5 & 37.6 & 179\,$-$\,229 & 205 & 37.9 & 2.9 &
1.1\,$-$\,2.0 & 1.7 \\
48.5 & 0.35 & 5.8 & 35.8 & 240\,$-$\,365 & 295 & 35.7 & 4.8 &
0.8\,$-$\,1.9 & 1.4 \\
\hline
\end{tabular}
%}
\end{center}\vspace{-0.4cm}
\tablefoot{(a): Value interval over the different observing sessions (three for 24\,GHz
and 12 for the rest of the frequencies).
(b): Averaged value for the final spectrum.
(c): Adding the two polarisations.}
%Note.-$\eta_{MB}$ and HPBW along the covered frequency range.
\end{table*}

Table\,\ref{table_telescope} shows the aperture efficiency ($\eta_{\rm A}$), the conversion factor between the flux ($S$) and antenna temperature ($T_{\rm A}^*$), 
the half power beam width (HPBW), the system temperature ($T_{\rm sys}$), the integration time, the noise root mean square (rms), and the source continuum emission in $T_{\rm A}^*$ along the covered frequency range.

\begin{figure}[h]
%\centering
\includegraphics[scale=0.65,angle=0]{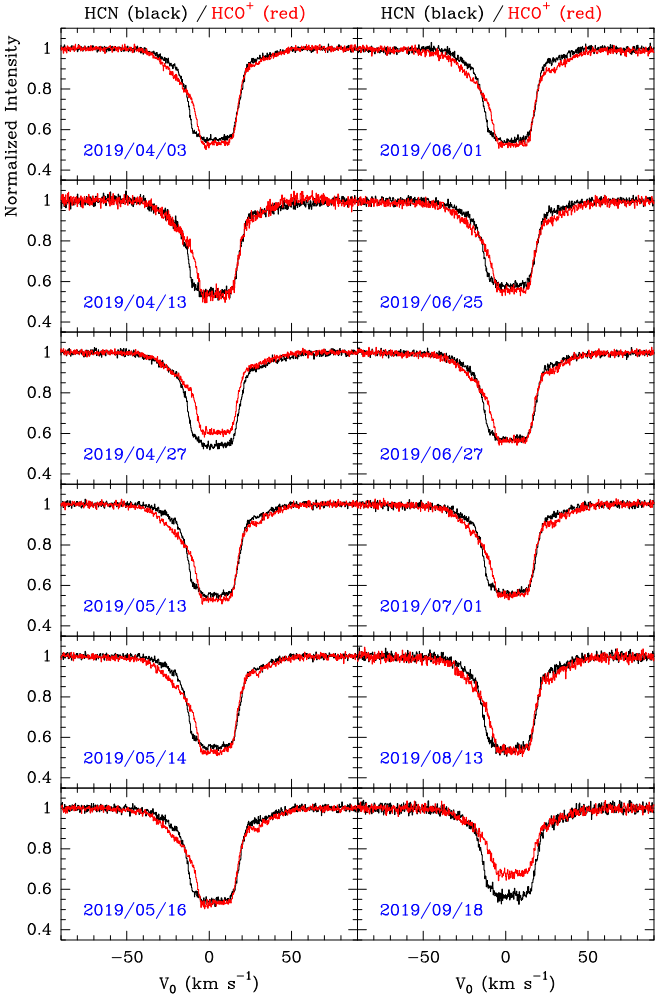}
\caption{Lines of HCO$^+$ ($J=1-0$) and HCN ($J=1-0$) observed as part of our data 
(see Sect.\,\ref{cont}).}
\label{fig_hcn_hco}
 \end{figure}

Figure\,\ref{fig_hcn_hco} shows the HCO$^+$ ($J=1-0$) and HCN ($J=1-0$) lines,
which are clearly saturated at the SW line of sight in our data.

%\clearpage

\section{Spectroscopic and observational line parameters}
Tables\,\ref{Table_CH3SH}\,$-$\,\ref{Table_NH2CN} show spectroscopic and
observational line parameters for the detected species.

%{\bf General notes to Appendix A tables}:
%(i) Frequencies, energy of the upper level of each transition ($E_{\rm u}/$k_{\rm B}$$ in K), Einstein coefficient for spontaneous emission ($A_{\rm ul}$ in s$^{-1}$), intrinsic line strength ($S_{\rm ul}$), and the level degeneracy ($g_{\rm u}$) from MADEX. The %velocity-integrated intensity"comosellameestoahora" ($\int \tau dv$ in units of 10$^{-3}$\,$\times$\,km\,s$^{-1}$), velocity ($v_{\rm 0}$), and FWHM line width ($\Delta$v) obtained by Gaussian fit.
%(ii) Parentheses indicate the uncertainty obtained by the Gaussian fitting programme.

\begin{table*} 
 \begin{center}
 \caption{Line parameters of CH$_3$SH.}  \label{Table_CH3SH}  
 \begin{tabular}{c c c r c c c c c c@{\vrule height 10pt depth 5pt width 0pt}}     % 11 columns 
 \hline\hline       
          
Transition &   $\nu_{\rm rest}$  & $\nu_{\rm z=0.885875}$  &  $E_{\rm u}$/$k_{\rm B}$ & $A_{\rm ul}$ &  $S_{\rm ul}\,\mu_{\rm a}^2$ &  $g_{\rm u}$ &  $\displaystyle{\int} \tau dv$  &  $v_{\rm 0}$ &  $v_{\rm FWHM}$  \rule[-0.3cm]{0cm}{0.8cm}\ \\ \cline{1-1} 

 $(J_{K})_{\rm u} \rightarrow (J_{K})_{\rm l}$ & [MHz] & [MHz] & [K] & [s$^{-1}$] & & & [10$^{-3}$\,km s$^{-1}$] & [km s$^{-1}$] & [km s$^{-1}$] \,\,\, \\   
            
              \hline
 A, 3$_{1}^{+}$ $\rightarrow$ 2$_{1}^{+}$  &    75085.910  & 39814.892 &  8.6 &  3.24 $\times$ 10$^{-6}$ & 4.60  &   7  & \,\,\,\,30\,(5)  &  $-$0.8\,(0.9)   &  11\,(2)     \\
 A, 3$_{0}^{+}$ $\rightarrow$ 2$_{0}^{+}$  &    75862.870  & 40226.881 &  5.1 &  3.75 $\times$ 10$^{-6}$ & 5.17  &   7  & \rdelim\}{2}{4mm}[ 85\,(8)$\dagger$]  &  \multirow{2}{*}{$-$0.7\,(0.5)}   &  \multirow{2}{*}{12\,(1)}     \\
 E, 3$_{0}$ $\rightarrow$ 2$_{0}$          &    75864.430  & 40227.709 &  5.2 &  3.76 $\times$ 10$^{-6}$ & 5.18  &   7  &    &    &        \\
% A, 3$_{2}^{-}$ $\rightarrow$ 2$_{2}^{-}$  &    75872.320  & 40231.892 & 19.5 &  2.09 $\times$ 10$^{-6}$ & 2.88  &   7  & \rdelim\}{2.0}{*}[ 9.0\,(3.3)\,\,]  &  \multirow{2.0}{*}{$-$0.4\,(0.8)}   &  \multirow{2.0}{*}{3.8\,(1.4)}      \\
% E, 3$_{-2}$ $\rightarrow$ 2$_{-2}$        &    75873.810  & 40232.682 & 17.3 &  2.09 $\times$ 10$^{-6}$ & 2.88  &   7  &    &     &       \\
 E, 3$_{1}$ $\rightarrow$ 2$_{1}$          &    75925.910  & 40260.309 &  8.4 &  3.35 $\times$ 10$^{-6}$ & 4.60  &   7  &  \,8\,(3)  &  $-$1.8\,(0.4)   &  2.4\,(0.7)      \\
 
\hline      
  \end{tabular}  
  \end{center}  
  \tablefoot{Spectroscopic data from \citet{Bettens_1999}. Parentheses indicate the uncertainty obtained by the Gaussian fitting programme.\\
  $\dagger$ Value obtained for the blended line.}
  \end{table*}

\begin{table*}
 \begin{center}
 \caption{Line parameters of l-C$_{3}$H$^+$.}  \label{Table_l-C3H+}  
 \begin{tabular}{c c c r c c c c c c@{\vrule height 10pt depth 5pt width 0pt}}     % 11 columns
 \hline\hline      
         
Transition &   $\nu_{\rm rest}$  & $\nu_{\rm z=0.885875}$  &  $E_{\rm u}$/$k_{\rm B}$ & $A_{\rm ul}$ &  $S_{\rm ul}$ &  $g_{\rm u}$ &  $\displaystyle{\int} \tau dv$  &  $v_{\rm 0}$ &  $v_{\rm FWHM}$  \rule[-0.3cm]{0cm}{0.8cm}\ \\ \cline{1-1}

 $J_{\rm u} \rightarrow J_{\rm l}$  & [MHz] & [MHz] & [K] & [s$^{-1}$] & & & [10$^{-3}$\,km s$^{-1}$] & [km s$^{-1}$] & [km s$^{-1}$] \,\,\, \\  
           
              \hline
  2 $\rightarrow$ 1  & 44979.544 & 23850.756  &   3.2  &  3.81 $\times$ 10$^{-6}$  &   2.0 &  5 &  57\,(8)   & $-$0.5\,(0.4)  & 5.4\,(0.8)\\            
  3 $\rightarrow$ 2  & 67468.856 & 35775.890  &   6.5  &  1.38 $\times$ 10$^{-5}$  &   3.0 &  7 &  84\,(10)   & $-$0.6\,(0.2) & 7.0\,(0.9) \\
  4 $\rightarrow$ 3  & 89957.617 & 47700.731  &  10.8  &  3.39 $\times$ 10$^{-5}$  &   4.0 &  9 &  55\,(6)    & $-$0.8\,(0.3) & 4.5\,(0.5) \\

 \hline      
  \end{tabular}  
  \end{center}  
  \tablefoot{Spectroscopic line parameters have been obtained using \texttt{MADEX} by fitting the rotational lines reported by \citet{Pety_2012} and \citet{Cuadrado2015}. Parentheses indicate the uncertainty obtained by the Gaussian fitting programme.}
  \end{table*}  

%  \,\,\,\,\,\,\,\,\,\,\,\,\,\,\,\,
  
    \begin{table*}
\begin{center}
\caption{Line parameters of C$_{3}$N.}  \label{Table_C3N}
\resizebox{1\textwidth}{!}{
\begin{tabular}{c c c c c c c l c c@{\vrule height 10pt depth 5pt width 0pt}}    
\hline\hline      

Transition &   $\nu_{\rm rest}$  & $\nu_{\rm z=0.885875}$  &  $E_{\rm u}$/$k_{\rm B}$ & $A_{\rm ul}$ &  $S_{\rm ul}$ &  $g_{\rm u}$ &  $\displaystyle{\int} \tau dv$  &  $v_{\rm 0}$ &  $v_{\rm FWHM}$  \rule[-0.3cm]{0cm}{0.8cm}\ \\ \cline{1-1}

$(N, J, F)_{\rm u} \rightarrow (N, J, F)_{\rm l}$ & [MHz] & [MHz] & [K] & [s$^{-1}$] & & & [10$^{-3}$ km s$^{-1}$] & [km s$^{-1}$] & [km s$^{-1}$] \,\,    \\

\hline                                                                                                      
${\mathrm{(6, 13/2, 13/2)\rightarrow(5, 11/2, 13/2)}}$ & 59361.053  & 31476.664  & 10.0  & 2.18 $\times$ 10$^{-7}$ & 0.15 & 14  &  \rdelim\}{6.5}{*}[55\,(5)$\dagger$] & \multirow{7.0}{*}{$-$0.5\,(0.2)} & \multirow{7.0}{*}{7.8\,(0.7)}   \\
${\mathrm{(6, 13/2, 11/2)\rightarrow(5, 11/2, 9/2)\,\,\,}}$  & 59361.383  & 31476.839  & 10.0  & 8.87 $\times$ 10$^{-6}$ & 5.38 & 12 &  &  &     \\
${\mathrm{(6, 13/2, 13/2)\rightarrow(5, 11/2, 11/2)}}$ & 59361.386  & 31476.840  & 10.0  & 8.91 $\times$ 10$^{-6}$ & 6.31 & 14 &  &  &     \\
${\mathrm{(6, 13/2, 15/2)\rightarrow(5, 11/2, 13/2)}}$ & 59361.418  & 31476.857  & 10.0  & 9.13 $\times$ 10$^{-6}$ & 7.38 & 16 &  &  &     \\
${\mathrm{(6, 13/2, 11/2)\rightarrow(5, 11/2, 11/2)}}$ & 59364.084  & 31478.271  & 10.0  & 2.53 $\times$ 10$^{-7}$ & 0.15 & 12 &  &  &     \\

${\mathrm{(6, 11/2, 11/2)\rightarrow(5, 9/2, 11/2)\,\,\,}}$  & 59377.528  & 31485.400  & 10.0  & 2.95 $\times$ 10$^{-7}$ & 0.18 & 12 & \rdelim\}{6.5}{*}[(a)] &  &     \\
${\mathrm{(6, 11/2, 9/2)\rightarrow(5, 9/2, 7/2)\,\,\,}}$    & 59380.159  & 31486.795  & 10.0  & 8.64 $\times$ 10$^{-6}$ & 4.36 & 10 &  &  &     \\
${\mathrm{(6, 11/2, 11/2)\rightarrow(5, 9/2, 9/2)\,\,\,\,\,\,}}$   & 59380.192  & 31486.812  & 10.0  & 8.70 $\times$ 10$^{-6}$ & 5.27 & 12 &  &  &     \\
${\mathrm{(6, 11/2, 13/2)\rightarrow(5, 9/2, 11/2)\,\,\,}}$  & 59380.194  & 31486.813  & 10.0  & 9.00 $\times$ 10$^{-6}$ & 6.36 & 14 &  &  &     \\
${\mathrm{(6, 11/2, 9/2)\rightarrow(5, 9/2, 9/2)\,\,\,}}$    & 59380.912  & 31487.194  & 10.0  & 3.56 $\times$ 10$^{-7}$ & 0.18 & 10 &  &  &     \\
                                                                             
${\mathrm{(7, 15/2, 15/2)\rightarrow(6, 13/2, 15/2)}}$  & 69255.887  & 36723.477  & 13.3 & 2.62 $\times$ 10$^{-7}$ & 0.13 & 16 & \rdelim\}{6.5}{*}[26\,(3)$\dagger$] & \multirow{7.0}{*}{$-$0.8\,(0.2)} & \multirow{7.0}{*}{4.2\,(0.5)}   \\
${\mathrm{(7, 15/2, 13/2)\rightarrow(6, 13/2, 11/2)}}$  & 69256.249  & 36723.669  & 13.3 & 1.44 $\times$ 10$^{-5}$ & 6.40 & 14 &  &  &     \\
${\mathrm{(7, 15/2, 15/2)\rightarrow(6, 13/2, 13/2)}}$  & 69256.252  & 36723.670  & 13.3 & 1.44 $\times$ 10$^{-5}$ & 7.33 & 16 &  &  &     \\
${\mathrm{(7, 15/2, 17/2)\rightarrow(6, 13/2, 15/2)}}$  & 69256.275  & 36723.683  & 13.3 & 1.47 $\times$ 10$^{-5}$ & 8.40 & 18 &  &  &     \\
${\mathrm{(7, 15/2, 13/2)\rightarrow(6, 13/2, 13/2)}}$  & 69258.947  & 36725.099  & 13.3 & 2.99 $\times$ 10$^{-7}$ & 0.13 & 14 &  &  &     \\

${\mathrm{(7, 13/2, 13/2)\rightarrow(6, 11/2, 13/2)}}$  & 69272.375  & 36732.220  & 13.3 & 3.41 $\times$ 10$^{-7}$ & 0.15 & 14 & \rdelim\}{6.5}{*}[20\,(3)$\dagger$]  & \multirow{7.0}{*}{$-$1.5\,(0.2)} & \multirow{7.0}{*}{3.9\,(0.7)}   \\
${\mathrm{(7, 13/2, 11/2)\rightarrow(6, 11/2, 9/2)\,\,\,}}$   & 69275.018  & 36733.621  & 13.3 & 1.41 $\times$ 10$^{-5}$ & 5.38 & 12 &  &  &    \\
${\mathrm{(7, 13/2, 13/2)\rightarrow(6, 11/2, 11/2)}}$  & 69275.042  & 36733.634  & 13.3 & 1.42 $\times$ 10$^{-5}$ & 6.31 & 14 &  &  &    \\
${\mathrm{(7, 13/2, 15/2)\rightarrow(6, 11/2, 13/2)}}$  & 69275.044  & 36733.635  & 13.3 & 1.45 $\times$ 10$^{-5}$ & 7.39 & 16 &  &  &    \\
${\mathrm{(7, 13/2, 11/2)\rightarrow(6, 11/2, 11/2)}}$  & 69275.738  & 36734.003  & 13.3 & 3.99 $\times$ 10$^{-7}$ & 0.15 & 12 &  &  &    \\

${\mathrm{(8, 17/2, 17/2)\rightarrow(7, 15/2, 17/2)}}$ & 79150.600  & 41970.226  & 17.1 & 3.07 $\times$ 10$^{-7}$ & 0.12 & 18 & \rdelim\}{6.5}{*}[(b)] &  &    \\
${\mathrm{(8, 17/2, 15/2)\rightarrow(7, 15/2, 13/2)}}$ & 79150.986  & 41970.431  & 17.1 & 2.17 $\times$ 10$^{-5}$ & 7.41 & 16 &  &  &     \\
${\mathrm{(8, 17/2, 17/2)\rightarrow(7, 15/2, 15/2)}}$ & 79150.988  & 41970.432  & 17.1 & 2.18 $\times$ 10$^{-5}$ & 8.35 & 18 &  &  &     \\
${\mathrm{(8, 17/2, 18/2)\rightarrow(7, 15/2, 17/2)}}$ & 79151.006  & 41970.441  & 17.1 & 2.21 $\times$ 10$^{-5}$ & 9.41 & 20 &  &  &     \\
${\mathrm{(8, 17/2, 15/2)\rightarrow(7, 15/2, 15/2)}}$ & 79153.682  & 41971.860  & 17.1 & 3.45 $\times$ 10$^{-7}$ & 0.12 & 16 &  &  &     \\

${\mathrm{(8, 15/2, 15/2)\rightarrow(7, 13/2, 15/2)}}$ & 79167.099  & 41978.975  & 17.1 & 3.87 $\times$ 10$^{-7}$ & 0.13 & 16 & \rdelim\}{6.5}{*}[19\,(4)$\dagger$] & \multirow{7.0}{*}{0.3\,(0.5)} & \multirow{7.0}{*}{5\,(1)}  \\
${\mathrm{(8, 15/2, 13/2)\rightarrow(7, 13/2, 11/2)}}$ & 79169.750  & 41980.380  & 17.1 & 2.15 $\times$ 10$^{-5}$ & 6.40 & 14 &  &  &     \\
${\mathrm{(8, 15/2, 15/2)\rightarrow(7, 13/2, 13/2)}}$ & 79169.768  & 41980.390  & 17.1 & 2.15 $\times$ 10$^{-5}$ & 7.33 & 16 &  &  &     \\
${\mathrm{(8, 15/2, 17/2)\rightarrow(7, 13/2, 15/2)}}$ & 79169.770  & 41980.391  & 17.1 & 2.19 $\times$ 10$^{-5}$ & 8.40 & 18 &  &  &     \\
${\mathrm{(8, 15/2, 13/2)\rightarrow(7, 13/2, 13/2)}}$ & 79170.446  & 41980.750  & 17.1 & 4.44 $\times$ 10$^{-7}$ & 0.13 & 14 &  &  &     \\
\hline    
\end{tabular}
}
\end{center}
\tablefoot{Spectroscopic data from CDMS catalogue. Parentheses indicate the uncertainty obtained by the Gaussian fitting programme.
%Rest frequencies, \mbox{$E_{\rm u}$}, \mbox{$A_{\rm ul}$}, \mbox{$S_{\rm ul}$}, and \mbox{$g_{\rm u}$} from CDMS catalogue. Parentheses indicate the uncertainty obtained by the Gaussian fitting programme.
Fully overlapping transitions are marked with connecting symbols. (a) Blended with A-CH$_3$CHO 3$_{1,2}$\,$\rightarrow$\,2$_{1,1}$. (b) Blended with A-CH$_3$CHO 4$_{1,3}$\,$\rightarrow$\,3$_{1,2}$.\\
$\dagger$ Value obtained for the blended line of the overlapping hyperfine components.}
\end{table*}

%TABLA HCOOH
\begin{table*}
\begin{center}
\caption{Line parameters of HCOOH.}\label{Table_HCOOH}  
\begin{tabular}{c c c r c c c c c c@{\vrule height 10pt depth 5pt width 0pt}}  
\hline\hline      
         
Transition &   $\nu_{\rm rest}$  & $\nu_{\rm z=0.885875}$  &  $E_{\rm u}$/$k_{\rm B}$ & $A_{\rm ul}$ &  $S_{\rm ul}$ &  $g_{\rm u}$ &  $\displaystyle{\int} \tau dv$  &  $v_{\rm 0}$ &  $v_{\rm FWHM}$  \rule[-0.3cm]{0cm}{0.8cm}\ \\ \cline{1-1}
     
$(J_{K_{\rm a},K_{\rm c}})_{\rm u} \rightarrow (J_{K_{\rm a},K_{\rm c}})_{\rm l}$  & [MHz] & [MHz] & [K] & [s$^{-1}$] & & & [10$^{-3}$\,km s$^{-1}$] & [km s$^{-1}$] & [km s$^{-1}$]  \,\,\, \\  
           
              \hline
3$_{1,3}$ $\rightarrow$ 2$_{1,2}$    & 64936.268   & 34432.965 &   9.4    &  2.45 $\times$ 10$^{-6}$  & 2.67  &  7  & 39\,(4)   & $-$1.0\,(0.2)  & 5.3\,(0.6)    \\
3$_{0,3}$ $\rightarrow$ 2$_{0,2}$    & 67291.121   & 35681.645 &   6.5    &  3.07 $\times$ 10$^{-6}$  & 3.00  &  7  & 34\,(3)   & $-$0.7\,(0.2)  & 4.2\,(0.4)    \\  
3$_{1,2}$ $\rightarrow$ 2$_{1,1}$    & 69851.954   & 37039.546 &   9.9    &  3.05 $\times$ 10$^{-6}$  & 2.67  &  7  & 42\,(2)   & $-$0.6\,(0.2)  & 6.1\,(0.5)    \\
4$_{1,4}$ $\rightarrow$ 3$_{1,3}$    & 86546.180   & 45891.790 &  13.6    &  6.35 $\times$ 10$^{-6}$  & 3.75  &  9  & 38\,(8)   & $-$0.6\,(0.7)  & 6\,(1)    \\
4$_{0,4}$ $\rightarrow$ 3$_{0,3}$    & 89579.168   & 47500.056 &  10.8    &  7.51 $\times$ 10$^{-6}$  & 4.00  &  9  & 27\,(6)   & $-$0.9\,(0.4)  & 2.8\,(0.6)     \\
\hline      
  \end{tabular}                                                                                                                                                          
  \end{center}  
  \tablefoot{Spectroscopic line parameters were obtained using \texttt{MADEX} by fitting all the rotational lines reported by \citet{Winnewisser_2002} and \citet{Cazzoli_2010b}. Parentheses indicate the uncertainty obtained by the Gaussian fitting programme.}
  \end{table*}

\begin{table*}
\begin{center}
\caption{Line parameters of CH$_2$CHCN.}  \label{Table_CH2CHCN}  
\begin{tabular}{c c c r c c c c c l@{\vrule height 10pt depth 5pt width 0pt}}     % 11 columns
\hline\hline      
         
Transition &   $\nu_{\rm rest}$  & $\nu_{\rm z=0.885875}$  &  $E_{\rm u}$/$k_{\rm B}$ & $A_{\rm ul}$ &  $S_{\rm ul}$ &  $g_{\rm u}$ &  $\displaystyle{\int} \tau dv$  &  $v_{\rm 0}$ &  $v_{\rm FWHM}$  \rule[-0.3cm]{0cm}{0.8cm}\ \\ \cline{1-1}

 $(J_{K_{\rm a},K_{\rm c}})_{\rm u} \rightarrow (J_{K_{\rm a},K_{\rm c}})_{\rm l}$ & [MHz] & [MHz] & [K] & [s$^{-1}$] & & & [10$^{-3}$\,km s$^{-1}$] & [km s$^{-1}$] & [km s$^{-1}$] \,\,\, \\  
       
              \hline
 7$_{1,7}$ $\rightarrow$ 6$_{1,6}$  &  64749.012  & 34333.671  &  14.6  &  2.11 $\times$ 10$^{-5}$ & 6.86  & 15  &  26.1\,(3.2)  & $-$0.2\,(0.4)  &  5.7\,(0.7)  \\
 7$_{0,7}$ $\rightarrow$ 6$_{0,6}$  &  66198.347  & 35102.193  &  12.7  &  2.30 $\times$ 10$^{-5}$ & 7.00  & 15  &  26.7\,(3.5)  & $-$0.6\,(0.3)  &  6.3\,(0.8)  \\
 7$_{1,6}$ $\rightarrow$ 6$_{1,5}$  &  67946.687  & 36029.263  &  15.2  &  2.44 $\times$ 10$^{-5}$ & 6.86  & 15  &  28.6\,(4.4)  & $-$0.9\,(0.4)  &  6.1\,(0.9)  \\
 8$_{1,8}$ $\rightarrow$ 7$_{1,7}$  &  73981.555  & 39229.299  &  18.2  &  3.19 $\times$ 10$^{-5}$ & 7.87  & 17  &  21.5\,(3.2)  & $-$2.2\,(0.3)  &  5.0\,(0.8)$\dagger$ \\
 8$_{0,8}$ $\rightarrow$ 7$_{0,7}$  &  75585.693  & 40079.906  &  16.4  &  3.45 $\times$ 10$^{-5}$ & 8.00  & 17  &  24.3\,(3.6)  & $-$0.2\,(0.4)  &  5.0\,(0.8)$\dagger$  \\
 8$_{2,6}$ $\rightarrow$ 7$_{2,5}$  &  76128.883  & 40367.937  &  25.1  &  3.31 $\times$ 10$^{-5}$ & 7.50  & 17  &   5.6\,(0.8)  & $-$1.5\,(0.7)  &  5.0\,(0.8)$\dagger$,$\dagger$$\dagger$ \\
 8$_{1,7}$ $\rightarrow$ 7$_{1,6}$  &  77633.824  & 41165.944  &  18.9  &  3.68 $\times$ 10$^{-5}$ & 7.87  & 17  &  16.8\,(2.9)  & $+$0.3\,(0.4)  &  5.0\,(0.8)$\dagger$,$\dagger$$\dagger$ \\

  \hline      
  \end{tabular}  
  \end{center}  
  \tablefoot{Spectroscopic line parameters were obtained using \texttt{MADEX} by fitting all the rotational lines reported by \citet{Kisiel_2009}. Parentheses indicate the uncertainty obtained by the Gaussian fitting programme.\\
  $\dagger$: Owing to the weakness of these lines, we fixed $v_{\rm FWHM}$ to 5.0\,km\,s$^{-1}$ in order to fit a Gaussian profile. We assume an uncertainty of 15\% for this value.\\
  $\dagger$$\dagger$: These lines are below a 3$\sigma$ level. We did not include them in the analysis of the rotational diagrams.}
  \end{table*}

   %TABLA o-H2CN   Tenemos (tienes) que enterarnos de cómo van los números cuánticos y escribirlo en la nota (no?).
 \begin{table*}
   \begin{center}
    \caption{Line parameters of o-H$_{2}$CN.}  \label{Table_o-H2CN}  
    \resizebox{1\textwidth}{!}{
   \begin{tabular}{c c c r c c c c c c@{\vrule height 10pt depth 5pt width 0pt}}     % 11 columns
   \hline\hline      
           
  Transition &   $\nu_{\rm rest}$  & $\nu_{\rm z=0.885875}$  &  $E_{\rm u}$/$k_{\rm B}$ & $A_{\rm ul}$ &  $S_{\rm ul}$ &  $g_{\rm u}$ &  $\displaystyle{\int} \tau dv$  &  $v_{\rm 0}$ &  $v_{\rm FWHM}$  \rule[-0.3cm]{0cm}{0.8cm}\ \\ \cline{1-1}
 
  $(N_{K_{\rm a},K_{\rm c}}, n, F)_{\rm u} \rightarrow (N_{K_{\rm a},K_{\rm c}}, n, F)_{\rm l}$  & [MHz] & [MHz] & [K] & [s$^{-1}$] & & & [10$^{-3}$\,km s$^{-1}$] & [km s$^{-1}$] & [km s$^{-1}$] \,\,\, \\    
               
                \hline  
         (1$_{0,1}$, 3, 5/2) $\rightarrow$ (0$_{0,0}$, 2, 3/2)  &  73345.486  & 38892.019 & 3.5 &  2.83 $\times$ 10$^{-6}$   &  0.57  &   6 &  15\,(4)   & $-$1.7\,(0.6)   &    4\,(1)  \\
         (1$_{0,1}$, 0, 7/2) $\rightarrow$ (0$_{0,0}$, 0, 5/2)  &  73349.648  & 38894.226 & 3.5 &  3.29 $\times$ 10$^{-6}$   &  0.89  &   8 &  20\,(6)   & $-$0.4\,(0.4)   &    4\,(1)  \\
         (1$_{0,1}$, 2, 5/2) $\rightarrow$ (0$_{0,0}$, 3, 3/2)  &  73409.042  & 38925.720 & 3.5 &  3.17 $\times$ 10$^{-6}$   &  0.64  &   6 &  22\,(6)   & $-$1.0\,(0.6)   &    4\,(1)  \\
         (1$_{0,1}$, 4, 5/2) $\rightarrow$ (0$_{0,0}$, 0, 5/2)  &  73444.240  & 38944.384 & 3.5 &  2.73 $\times$ 10$^{-6}$   &  0.55  &   6 &  15\,(4)   & $-$0.4\,(0.8)   &    6\,(2)  \\
 \hline      
  \end{tabular}  
  }
  \end{center}  
  \tablefoot{Spectroscopic data from CDMS catalogue. Parentheses indicate the uncertainty obtained by the Gaussian fitting programme.}
  %Rest frequencies, \mbox{$E_{\rm u}$}, \mbox{$A_{\rm ul}$}, \mbox{$S_{\rm ul}$}, and \mbox{$g_{\rm u}$} from CDMS catalogue. Parentheses indicate the uncertainty obtained by the Gaussian fitting programme.}
  \end{table*}

\begin{table*}
 \begin{center}
 \caption{Line parameters of OCS.}  \label{Table_OCS}  
 \begin{tabular}{c c c r c c c c c c@{\vrule height 10pt depth 5pt width 0pt}}    
 \hline\hline      
             
  Transition &   $\nu_{\rm rest}$  & $\nu_{\rm z=0.885875}$  &  $E_{\rm u}$/$k_{\rm B}$ & $A_{\rm ul}$ &  $S_{\rm ul}$ &  $g_{\rm u}$ &  $\displaystyle{\int} \tau dv$  &  $v_{\rm 0}$ &  $v_{\rm FWHM}$  \rule[-0.3cm]{0cm}{0.8cm}\ \\ \cline{1-1}       
  $J_{\rm u} \rightarrow J_{\rm l}$ & [MHz] & [MHz] & [K] & [s$^{-1}$] &  &  & [10$^{-3}$\,km s$^{-1}$] & [km s$^{-1}$] & [km s$^{-1}$] \,\,\,  \\           
 \hline
     5 $\rightarrow$ 4   &  60814.270  & 32247.244 &   8.8 & 6.09 $\times$ 10$^{-7}$ &  5.0 &  11  & 52\,(4) & $-$0.8\,(0.1) &  5.4\,(0.4)  \\
     6 $\rightarrow$ 5   &  72976.781  & 38696.510 &  12.3 & 1.07 $\times$ 10$^{-6}$ &  6.0 &  13  & 66\,(4) & $-$0.8\,(0.1) &  5.8\,(0.4)  \\
     7 $\rightarrow$ 6   &  85139.104  & 45145.677 &  16.3 & 1.71 $\times$ 10$^{-6}$ &  7.0 &  15  & 42\,(10) & $-$0.7\,(0.8) &  6\,(1)  \\
\hline      
\end{tabular}  
 \tablefoot{Spectroscopic line parameters were obtained using \texttt{MADEX} by fitting all the rotational lines reported by \citet{Golubiatnikov_2005}. Parentheses indicate the uncertainty obtained by the Gaussian fitting programme.}                              
\end{center}
\end{table*}

 \begin{table*}
\begin{center}
\caption{Line parameters of SO$_{2}$.}  \label{Table_SO2}  
\begin{tabular}{c c c r c r c c c c@{\vrule height 10pt depth 5pt width 0pt}}    
\hline\hline      
     
Transition &   $\nu_{\rm rest}$  & $\nu_{\rm z=0.885875}$  &  $E_{\rm u}$/$k_{\rm B}$ & $A_{\rm ul}$ &  $S_{\rm ul}$ &  $g_{\rm u}$ &  $\displaystyle{\int} \tau dv$  &  $v_{\rm 0}$ &  $v_{\rm FWHM}$  \rule[-0.3cm]{0cm}{0.8cm}\ \\ \cline{1-1}
   
$(J_{K_{\rm a},K_{\rm c}})_{\rm u} \rightarrow (J_{K_{\rm a},K_{\rm c}})_{\rm l}$  & [MHz] & [MHz] & [K] & [s$^{-1}$] &  &  & [10$^{-3}$\,km s$^{-1}$] & [km s$^{-1}$] & [km s$^{-1}$]  \,\,\,  \\
         
\hline
   4$_{1,3}$ $\rightarrow$ 4$_{0,4}$ & 59224.870 & 31404.452  &  12.0 & 3.01 $\times$ 10$^{-6}$ & 4.20 &  9 & 86\,(3) & $-$0.9\,(0.1) & 6.8\,(0.3)   \\    
   6$_{1,5}$ $\rightarrow$ 6$_{0,6}$ & 68972.159 & 36573.028  &  22.5 & 4.34 $\times$ 10$^{-6}$ & 5.54 & 13 & 49\,(5) & $-$0.7\,(0.3) & 8.3\,(0.8)  \\    
   1$_{1,1}$ $\rightarrow$ 0$_{0,0}$ & 69575.929 & 36893.182  &   3.3 & 3.49 $\times$ 10$^{-6}$ & 1.00 &  3 & 90\,(5) & $-$1.0\,(0.1) & 6.7\,(0.4)  \\    
  \hline                                                                    
\end{tabular}                                                                
\end{center}                                                                  
\tablefoot{Spectroscopic line parameters were obtained using \texttt{MADEX} by fitting all the rotational lines reported by \citet{Muller_2000c}. Parentheses indicate the uncertainty obtained by the Gaussian fitting programme.}                                                            
\end{table*}

 \begin{table*}
 \begin{center}
 \caption{Line parameters of NH$_2$CN.}  \label{Table_NH2CN}
 \begin{tabular}{c c c r c c c c c l@{\vrule height 10pt depth 5pt width 0pt}}     % 11 columns
 \hline\hline

Transition &   $\nu_{\rm rest}$  & $\nu_{\rm z=0.885875}$  &  $E_{\rm u}$/$k_{\rm B}$ & $A_{\rm ul}$ &  $S_{\rm ul}$ &  $g_{\rm u}$ & $\displaystyle{\int} \tau dv$  &  $v_{\rm 0}$ &  $v_{\rm FWHM}$ \rule[-0.3cm]{0cm}{0.8cm}\ \\ \cline{1-1}

 $(J_{K_{\rm a},K_{\rm c}})_{\rm u} \rightarrow (J_{K_{\rm a},K_{\rm c}})_{\rm l}$ & [MHz] & [MHz] & [K] & [s$^{-1}$] & & & [10$^{-3}$\,km s$^{-1}$] & [km s$^{-1}$] & [km s$^{-1}$] \,\,\, \\

\hline
O$^+$, 3$_{1,3}$ $\rightarrow$ 2$_{1,2}$    & 59587.700  & 31596.845  & 20.2   &  1.76 $\times$ 10$^{-5}$  &   8.00 &  21    &  30\,(3)   & $-$0.7\,(0.3)   & 7.2\,(0.7)  \\
O$^+$, 3$_{0,3}$ $\rightarrow$ 2$_{0,2}$    & 59985.700  & 31807.888  & 5.8   &  2.02 $\times$ 10$^{-5}$  &   3.00 &  7     &  16\,(3)   & $-$1.4\,(0.7)   & 7.0\,(1.0)$\dagger$  \\
O$^+$, 3$_{1,2}$ $\rightarrow$ 2$_{1,1}$    & 60378.911  & 32016.391  & 20.3   &  1.83 $\times$ 10$^{-5}$  &   8.00 &  21    &  34\,(4)   & $-$0.4\,(0.3)   & 7.4\,(0.9)  \\
O$^+$, 4$_{1,4}$ $\rightarrow$ 3$_{1,3}$    & 79449.729  & 42128.842  & 24.0   &  4.55 $\times$ 10$^{-5}$  &  11.25 &  27    &  43\,(5)   & $-$1.2\,(0.3)   & 6.0\,(0.7)  \\
O$^+$, 4$_{0,4}$ $\rightarrow$ 3$_{0,3}$    & 79979.596  & 42409.808  & 9.6   &  4.95 $\times$ 10$^{-5}$  &   4.00 &   9    &  30\,(4)   & $-$1.4\,(1.2)   & 7.0\,(1.0)$\dagger$  \\
O$^+$, 4$_{1,3}$ $\rightarrow$ 3$_{1,2}$    & 80504.600  & 42688.195  & 24.2   &  4.74 $\times$ 10$^{-5}$  &  11.25 &  27    &  48\,(6)   & $-$0.4\,(0.3)   & 6.6\,(0.9)  \\

 \hline
  \end{tabular}
  \end{center}
  \tablefoot{Spectroscopic data from JPL catalogue (\texttt{https://spec.jpl.nasa.gov/}). Parentheses indicate the uncertainty
obtained by the Gaussian fitting programme. \\
$\dagger$: Owing to the weakness of these lines, we fixed $v_{\rm FWHM}$ to 7.0\,km\,s$^{-1}$ in order to fit a Gaussian profile. We assume an uncertainty of 15\% for this value.}
  \end{table*}

\section{Rotational diagrams}
\label{appendix_dr}

\begin{figure*}[h!]
%\centering
%\vspace{1cm}
%\vspace{0.3cm}
\includegraphics[scale=0.4, angle=0]{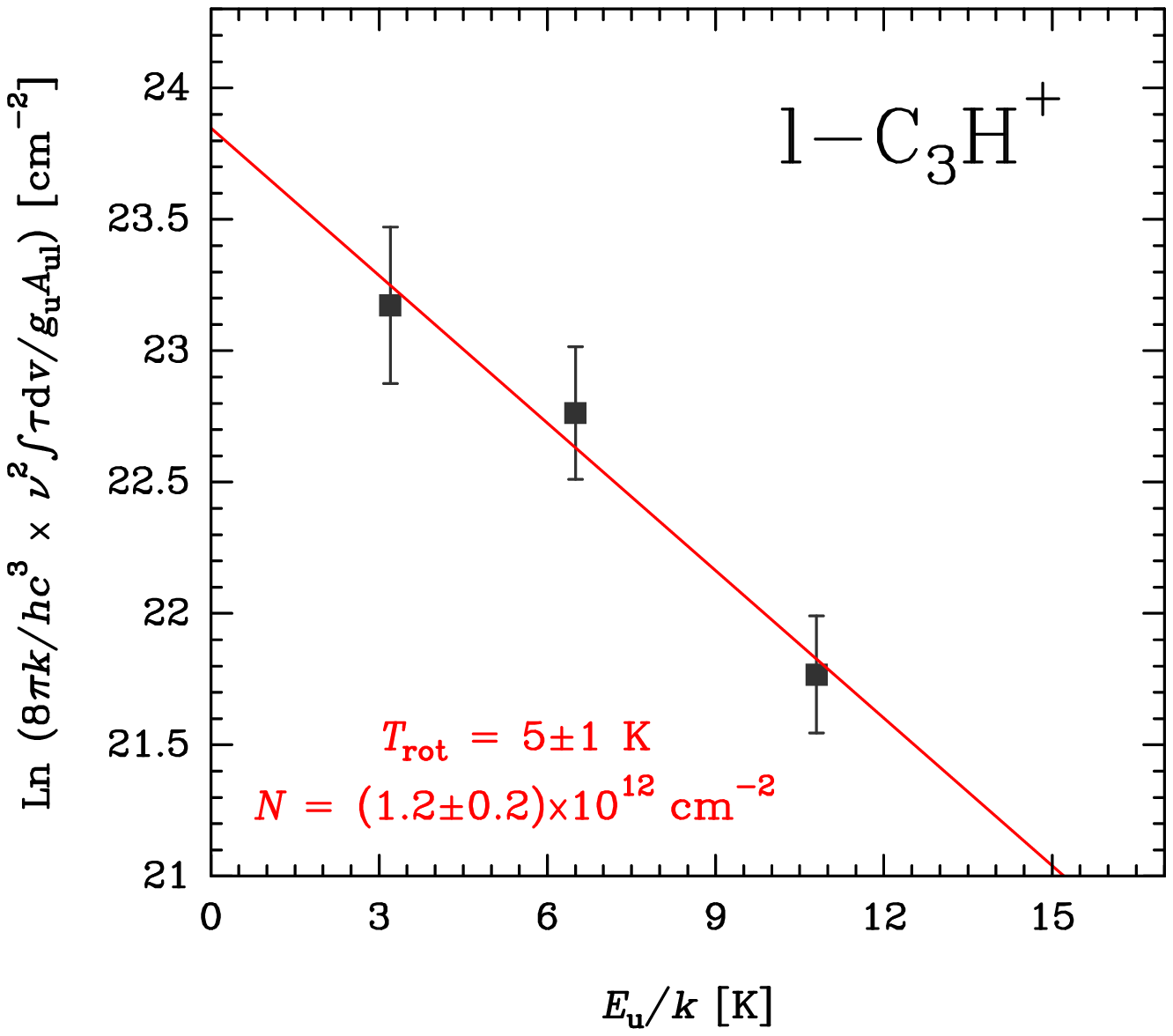} \hspace{0.5cm}
\includegraphics[scale=0.4, angle=0]{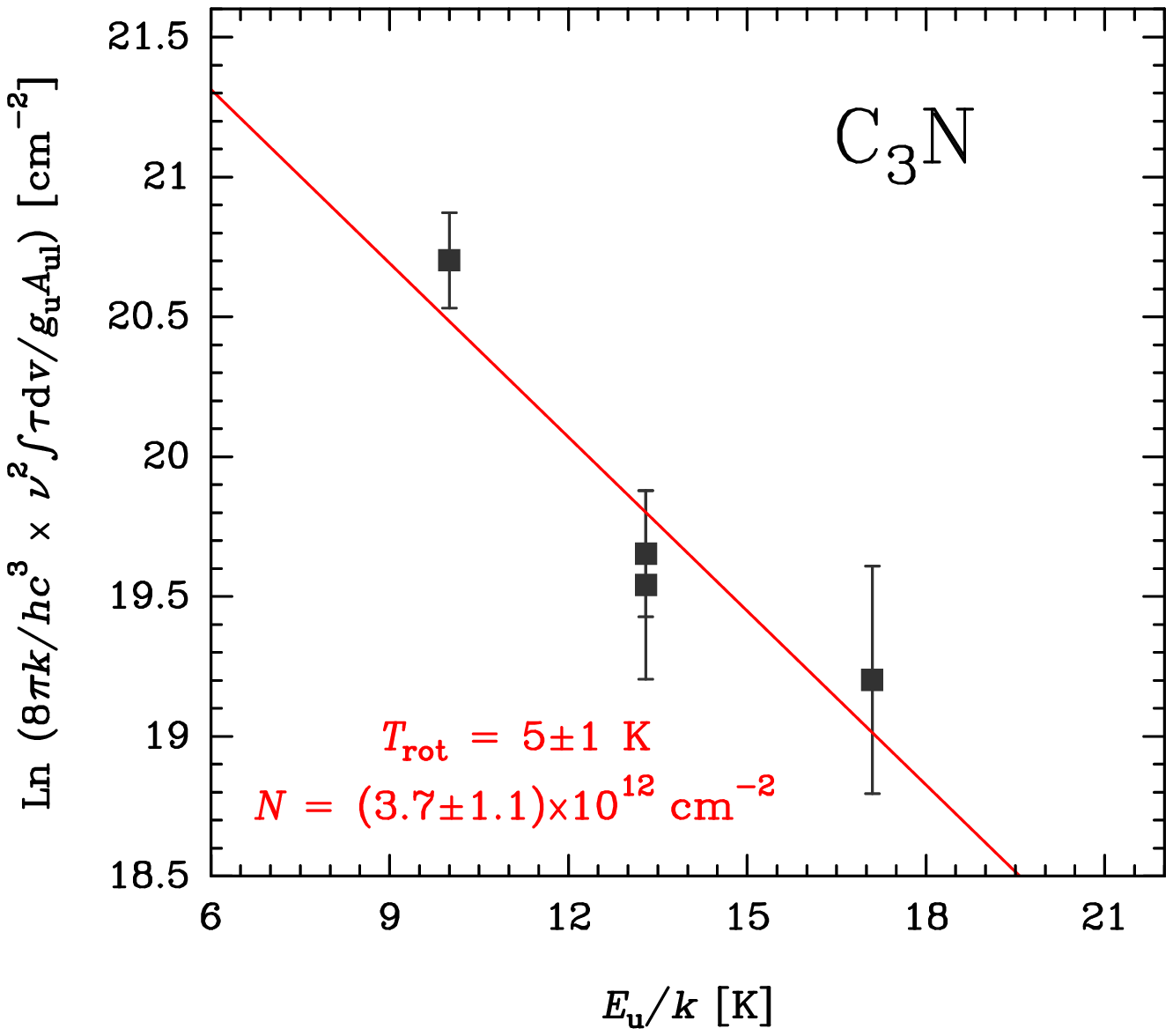} \hspace{0.5cm} 
\vspace{0.5cm}
\\
\vspace{0.5cm}
\includegraphics[scale=0.4, angle=0]{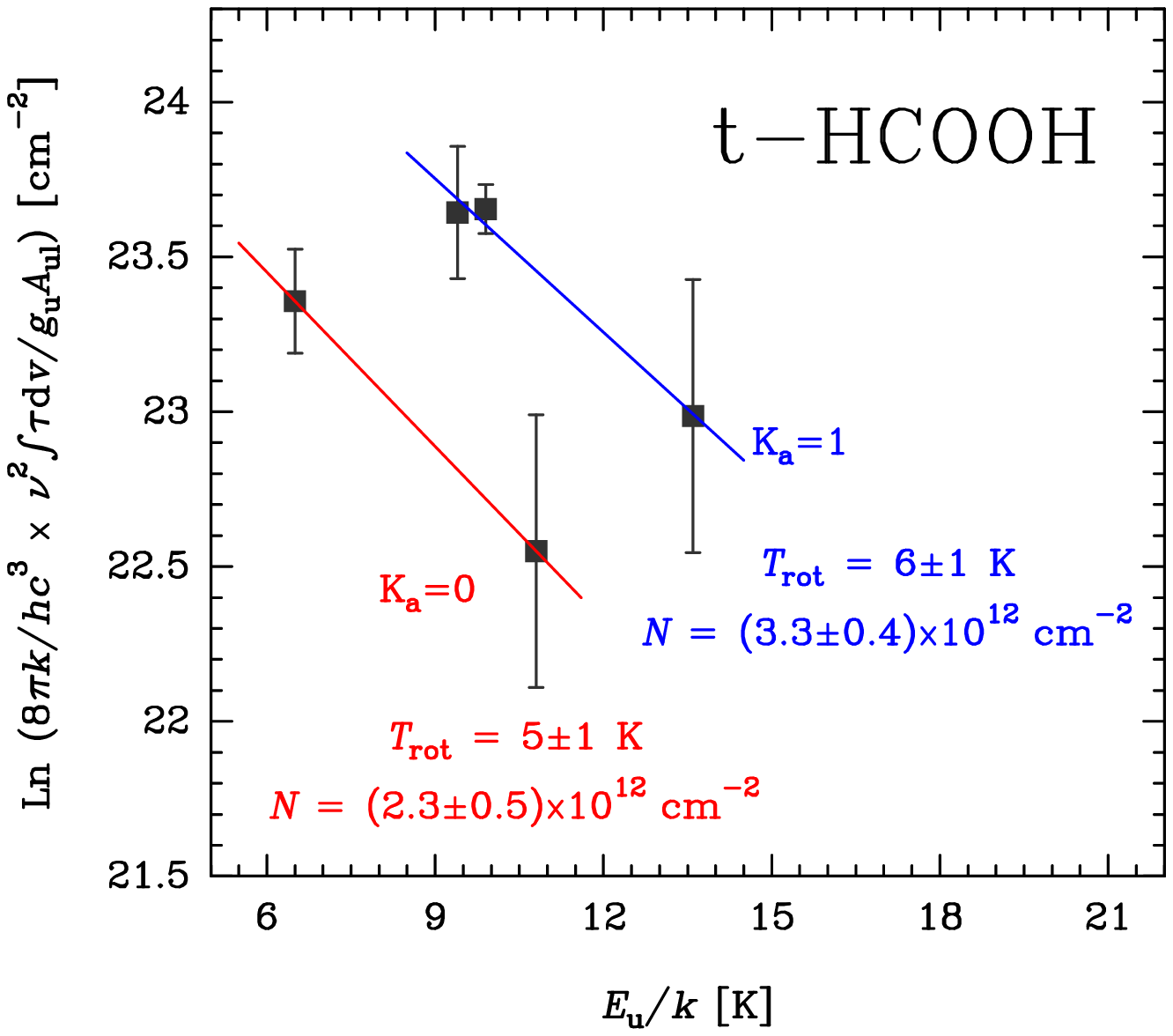} \hspace{0.5cm}
\includegraphics[scale=0.4, angle=0]{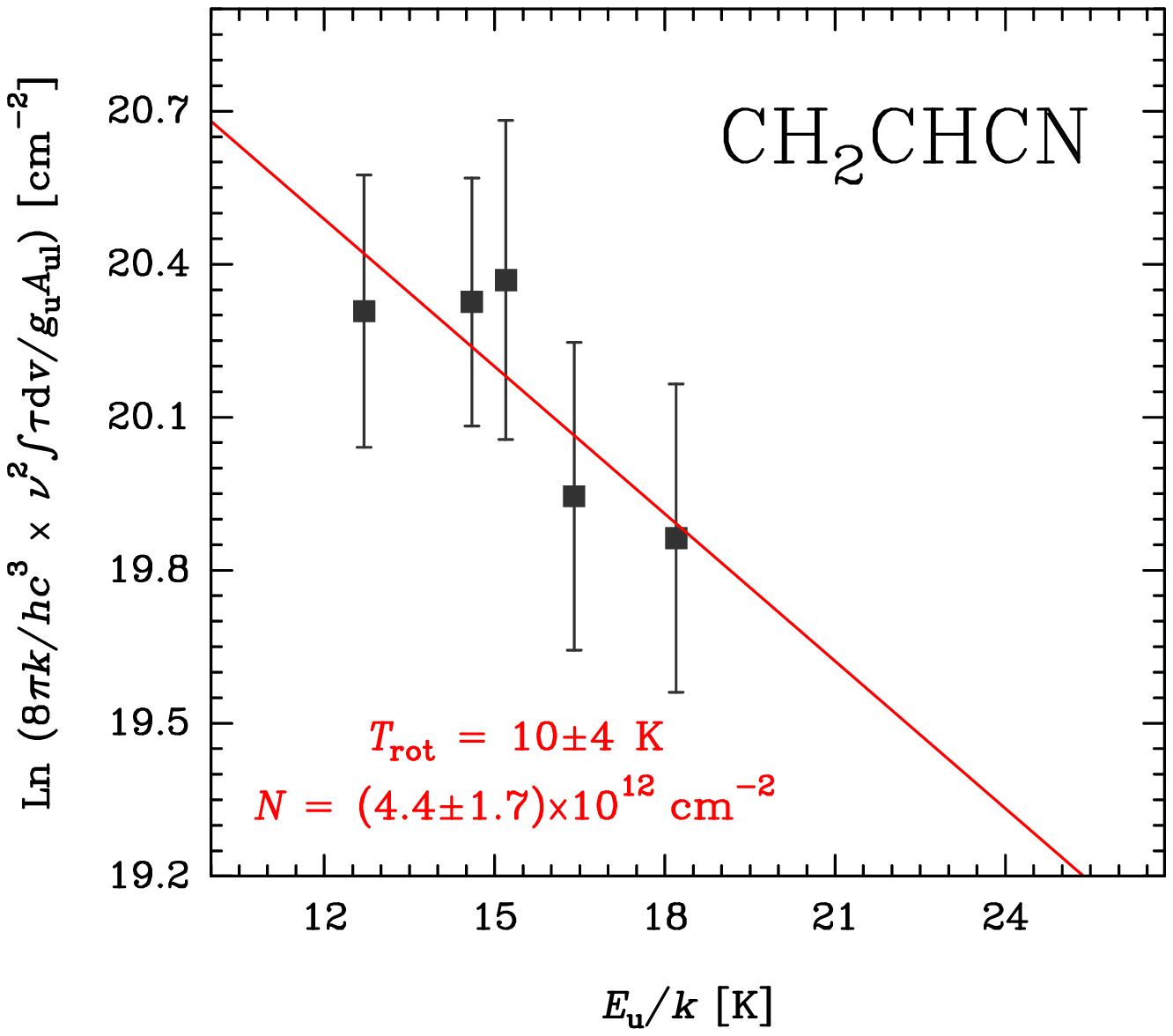} \hspace{0.5cm} \\
%\vspace{1cm}
\includegraphics[scale=0.4, angle=0]{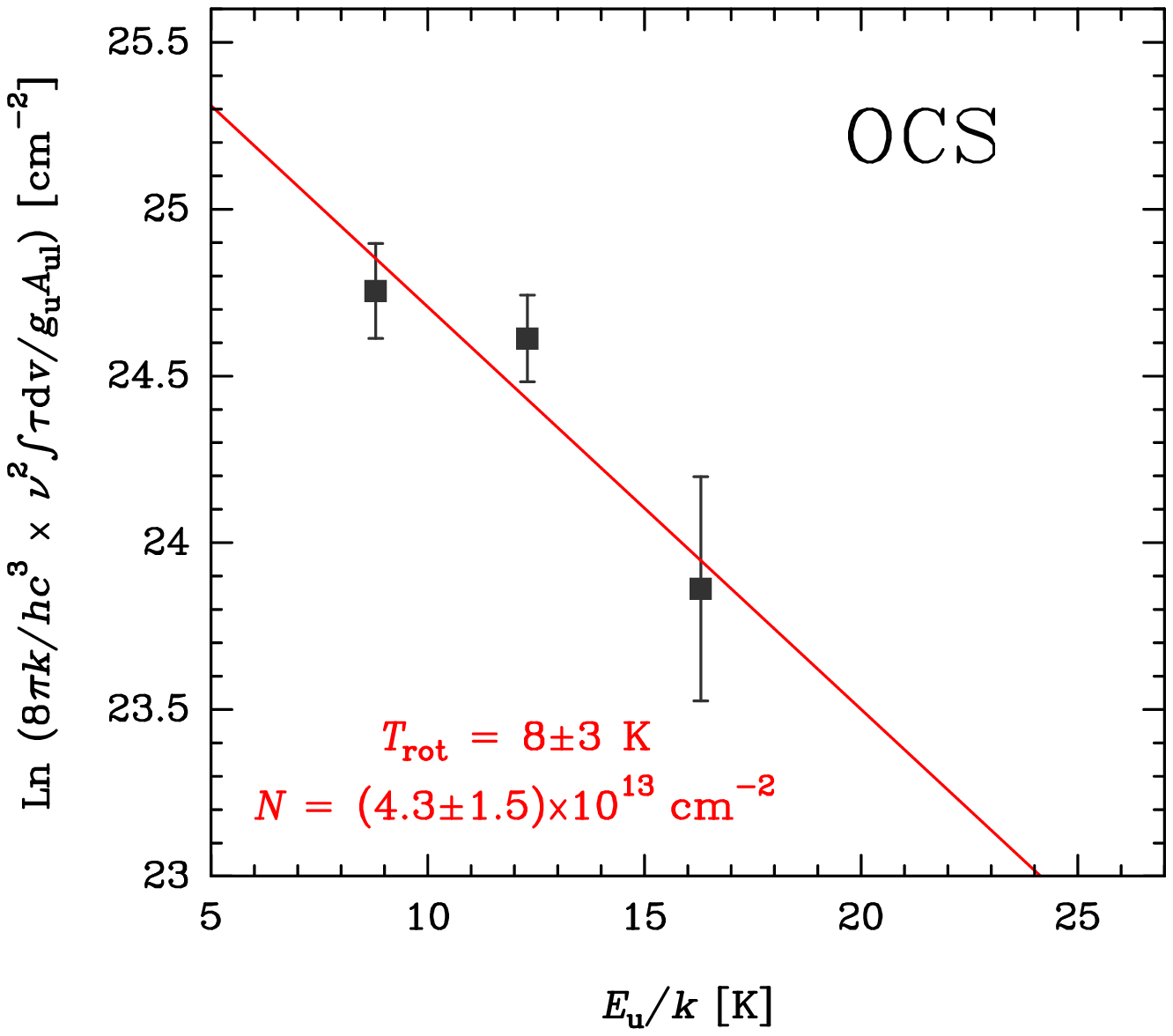} \hspace{0.5cm}
\includegraphics[scale=0.4, angle=0]{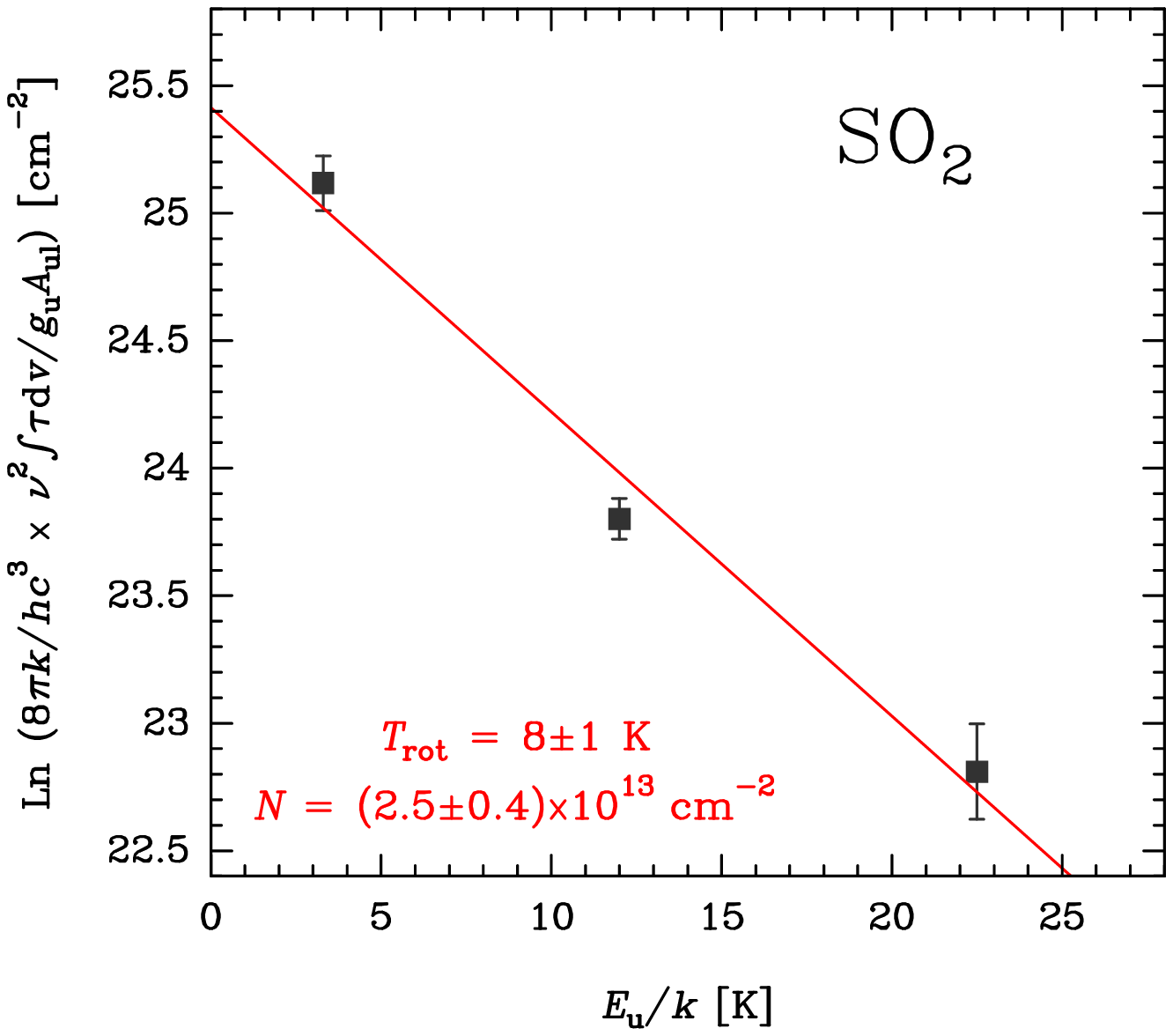}  \hspace{0.5cm}
\includegraphics[scale=0.4, angle=0]{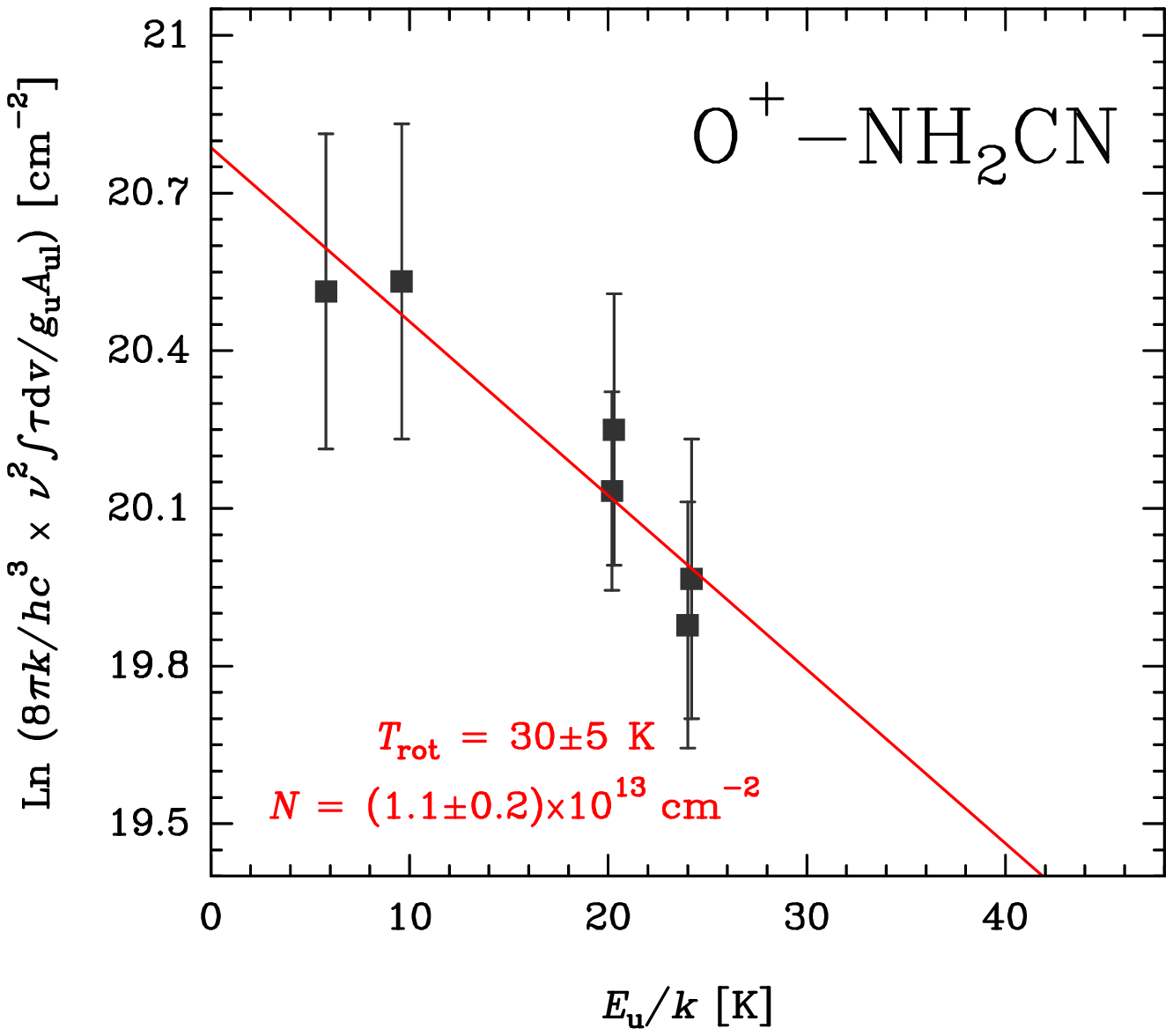}  \\
\caption{Rotational diagrams of the detected molecules towards PKS\,1830$-$211.
Derived values of the rotational temperature, $T_{\rm rot}$, column density, $N$, and their respective uncertainties are
also indicated for each molecule.}\label{fig_DR}
\end{figure*}

As we detected several molecules in more than one transition,
we can estimate rotational temperatures ($T_{\rm rot}$)
and molecular column densities ($N$) for the detected species by constructing
rotational diagrams (see e.g. \citealt{Goldsmith1999}).
This analysis assumes the Rayleigh-Jeans approximation, optically thin lines,
and LTE conditions. 
The equation that derives the total column density under these conditions
can be re-arranged as

\begin{equation}
{\rm \ln} \left(\frac{8 \pi k_{\rm B} \nu^2 \int{\tau_{\rm \nu} dv}}{h c^3 A_{\rm ul} g_{\rm u}}\right) = {\rm \ln} \left(\frac{N}{Q_{\rm rot} T_{\rm rot}}\right) - \frac{E_{\rm u}}{k_{\rm B} T_{\rm rot}}
\label{eq_RD}
,\end{equation}
where $g_u$ is the statistical weight in the upper level,
$A_{\rm ul}$ is the Einstein A-coefficient for
spontaneous emission, $Q_{\rm rot}$ is the rotational partition
function which depends on $T_{\rm rot}$,
$E_{\rm u}$ is the upper level energy, and $\nu$ is the frequency
of the transition. We obtained the integrated opacity of the line, $\int{\tau_{\rm \nu} dv}$, by multiplying
$\tau_{\rm \nu}$ by $v_{\rm FWHM}$.

The first term of Eq.\,\ref{eq_RD}, which only depends on spectroscopic and
observational line parameters, is plotted as a function of $E_{\rm u}$/$k_{\rm B}$
for the different detected lines. Thus,
the $T_{\rm rot}$ and $N$ can be derived
by performing a linear least square fit to the points (see Fig.\,\ref{fig_DR}).
Although we detected four lines of H$_2$CN, it is not possible to
perform a rotational diagram for this species since these lines correspond
to different hyperfine structure transitions with equal 
($N_{K_{\rm a},K_{\rm c}}$)$_{\rm u}$\,$\rightarrow$\,($N_{K_{\rm a},K_{\rm c}}$)$_{\rm l}$
state (and also equal
$E_{\rm u}$/$k_{\rm B}$, see Table\,\ref{Table_o-H2CN}).

Under the conditions of a subthermally excited gas, the rotational population diagram
of symmetric- and asymmetric-top molecules such as HCOOH
shows separate rotational ladders for each set of transitions with the same $K_{\rm a}$
quantum number. For very polar molecules,
subthermal excitation happens at relatively high densities.
Therefore, accurate column densities and
rotational temperatures from rotational diagrams can only be
obtained if the individual column densities for each rotational
ladder are computed independently (see \citealt{Cuadrado2016,Cuadrado2017}).
This effect has also prevented the analysis of the CH$_3$SH lines
since we have only detected one line with $K_{\rm a}$\,=\,0
and another with $K_{\rm a}$\,=\,1 per state (A/E).

For HCOOH, we built specific rotational diagrams for
different sets of lines with the same $K_{\rm a}$ quantum number.
The total column density of the molecule is obtained
by adding the column density of each rotational ladder. 
CH$_2$CHCN shows a similar tendency. However, due to the large uncertainties
of the individual points and the limited number of observed transitions,
we performed the rotational diagram from a single linear least square fitting.
For the other molecules, it is possible
to fit the lines from different $K$-ladders
by a single $T_{\rm rot}$ and $N$.

Due to the reduced number of rotational transitions, 
the rotational diagram of O$^+$-NH$_2$CN
(the lower inversion state of NH$_2$CN)
was built by taking the different
statistical weights for the $ortho$ ($K_{\rm a}$\,=\,1) and $para$ ($K_{\rm a}$\,=\,0) states into account (see $g_{\rm u}$ in Table\,\ref{Table_NH2CN}).
Interestingly, the rotational diagram of the very polar and asymmetric-top NH$_2$CN shows the points
of different $K_{\rm a}$ ladders merging in a single straight line. 
This characteristic straight diagram is typically seen
towards high density regions, such as hot cores (see e.g. \citealt{Lopez2014}).
For this kind of species, this merging of the $K_{\rm a}$ ladders only 
occurs at very high gas densities, which are higher
than the critical density for collisional excitation.

For C$_3$N, different hyperfine structure components of the same ($N$,\,$F$)$_{\rm u}$\,$\rightarrow$\,($N$,\,$F$)$_{\rm l}$ transition are blended in a single line.
Thus, to correctly determine $T_{\rm rot}$ and $N$,  
the line strength ($S_{\rm ul}$) was calculated as the sum of all allowed hyperfine
components of each ($N$,\,$F$)$_{\rm u}$\,$\rightarrow$\,($N$,\,$F$)$_{\rm l}$ transition.
The characteristic frequency ($\nu$) was determined using the weighted average with the
relative strength of each line as weight and $A_{\rm ul}$
was calculated using these derived values.
%and the usual
%relation.

Results for $T_{\rm rot}$ and $N$ using the population diagram procedure are shown in Table\,\ref{Table_dipole_moments} and Fig.\,\ref{fig_DR}.
The uncertainties were calculated using the statistical
errors given by the linear least squares fit for the slope and the
intercept. The individual errors of the data points are
those derived by taking into account the uncertainty obtained in the determination 
of the observed line parameters (see Tables\,\ref{Table_CH3SH}\,$-$\,\ref{Table_NH2CN}).

\end{appendix}

%\section{}    

\end{document}